\documentclass[final,3p,times,authoryear]{elsarticle}

\usepackage{amssymb}

\usepackage{natbib,twoopt}
\usepackage[breaklinks=false]{hyperref}  
\bibpunct{(}{)}{;}{a}{}{,}

\usepackage{color}
\usepackage{txfonts}
\usepackage{url}

\journal{Icarus}

\begin{document}

\begin{frontmatter}



\title{Haumea's thermal emission revisited in the light of the occultation results}


\author[MPE]{T. M\"{u}ller}
\author[KO]{Cs. Kiss}
\author[MPE]{V. Al\'{i}-Lagoa}
\author[IAA]{J. L. Ortiz}
\author[CNRS]{E. Lellouch}
\author[IAA]{P. Santos-Sanz}
\author[CNRS]{S. Fornasier}
\author[KO]{G. Marton}
\author[LO]{M. Mommert}
\author[KO,BU]{A. Farkas-Tak\'{a}cs}
\author[LO]{A. Thirouin}
\author[MPSS]{E. Vilenius}

\address[MPE]{Max-Planck-Institut f{\"u}r extraterrestrische Physik, Giessenbachstrasse 1, 85748 Garching, Germany}
\address[KO]{Konkoly Observatory, Research Centre for Astronomy and Earth Sciences, Hungarian Academy of Sciences, H-1121 Budapest, Konkoly Thege Mikl{\'o}s {\'u}t 15-17, Hungary}
\address[IAA]{Departamento de Sistema Solar, Instituto de Astrof{\'i}sica de Andaluc{\'i}a (CSIC), Glorieta de la Astronom{\'i}a s/n, 18008 Granada, Spain}
\address[CNRS]{LESIA-Observatoire de Paris, CNRS, UPMC Univ. Paris 06, Univ. Paris-Diderot, France}
\address[LO]{Lowell Observatory, 1400 W Mars Hill Rd, 86001, Flagstaff, Arizona, USA}
\address[MPSS]{Max-Planck-Institut f{\"u}r Sonnensystemforschung, Justus-von-Liebig-Weg 3, 37077 G{\"o}ttingen, Germany}
\address[BU]{E{\"o}tv{\"o}s Lor{\'a}nd University, P{\'a}zm{\'a}ny P{\'e}ter s. 1/A, H-1171 Budapest, Hungary}

\begin{abstract}
A recent multi-chord occultation measurement of the dwarf planet (136108) Haumea
\citep{Ortiz2017} revealed an elongated shape with the longest
axis comparable to Pluto's mean diameter. The chords also indicate a ring
around Haumea's equatorial plane,  where its largest moon, Hi'iaka, is also
located. The Haumea occultation size estimate (size of an equal-volume
sphere\footnote{D$_{equ}$ = 2$\cdot$ (a$\cdot$b$\cdot$c)$^{1/3}$.} D$_{equ}$ = 1595\,km)
is larger than previous radiometric solutions
(equivalent sizes in the range between 1150 and 1350\,km),
which lowers the object's density to about 1.8\,g/cm$^3$, a value closer
to the densities of other large TNOs. We present unpublished and also reprocessed
Herschel and Spitzer mid- and far-infrared measurements. We compare 100- and
160-$\mu$m thermal lightcurve amplitudes - originating from Haumea itself - with models of
the total measured system fluxes (ring, satellite, Haumea) from 24 -- 350\,$\mu$m. The combination with
results derived from the occultation measurements allows us to reinterpret the
object's thermal emission. Our radiometric studies show that Haumea's crystalline water ice
surface must have a thermal inertia of about 5\,J K$^{-1}$ m$^{-2}$s$^{-1/2}$
(combined with a root mean square of the surface slopes of 0.2).
We also have indications that the satellites (at least Hi'iaka) must have high geometric
albedos $\gtrsim$0.5, otherwise the derived thermal amplitude would be inconsistent with the total
measured system fluxes at 24, 70, 100, 160, 250, and 350\,$\mu$m. The
high albedos imply sizes of about 300 and 150\,km for Hi'iaka and Namaka,
respectively, indicating unexpectedly high densities $>$1.0\,g\,cm$^{-3}$ for TNOs this small,
and the assumed collisional formation from Haumea's icy crust.
We also estimated the thermal emission of the ring for the time period 1980-2030,
showing that the contribution during the Spitzer and Herschel epochs was
small, but not negligible.
Due to the progressive opening of the ring plane, the ring emission will
be increasing in the next decade when JWST is operational.
In the MIRI 25.5\,$\mu$m band it will also be possible
to obtain a very high-quality thermal lightcurve to test the derived Haumea properties.
\end{abstract}

\begin{keyword}



\end{keyword}

\end{frontmatter}


\section{Introduction}
\label{sec:intro}

Haumea is a large trans-Neptunian object (TNO) discovered in 2003, with
pre-discovery observations going back to Palomar Mountain DSS\footnote{Digitized Sky Survey} images from 1955. Its
very blue color \citep{Tegler2007} and a surface covered with crystalline
water ice \citep{Merlin2007,Tegler2007,Trujillo2007} makes
it very unique among the large TNOs.
It is known to have two satellites \citep{Brown2005,Brown2006} and a ring \citep{Ortiz2017}.
Both satellites show signs of water ice on the surface
\citep{Barkume2006, Fraser2009}, and the water-ice absorption
features are at least as deep as Haumea's. These moons are thought
to have been formed either by a catastrophic impact that excavated
them from the proto-Haumea ice mantle \citep{Brown2007} or from rotational
fission \citep{Ortiz2012}. \citet{Brown2007} reported that several other
dynamically clustered objects show similar water ice absorption features
and proposed that they all originated in an impact event at least 1 Gyr ago.
\citet{Vilenius2018} studied a range of
Haumea family members which were seen by Spitzer and Herschel. They determined Haumea-like
high albedos (average 0.48) and a distinct slope of the cumulative size distribution,
different from dynamical interlopers.
\citet{Lacerda2008} and \citet{Lacerda2009}
performed multi-color lightcurves and found indications for surface
heterogeneity on Haumea, some parts appearing to be redder and darker
than the rest. However, \citet{Pinilla-Alonso2009} did not see any significant
variations in their spectra taken at different rotational phases. More sensitive
NIR spectroscopy measurements \citep{Gourgeot2016} confirmed the presence of
crystalline water ice over large parts of the surface and also showed that the
so-called dark red spot (DRS) region was slightly redder spectrally, yet, still showing
the crystalline ice signature. 

Based on a large visible lightcurve amplitude, a fast rotation, and assuming
hydrostatic equilibrium, \citet{Rabinowitz2006} estimated the dimentions of
980\,km $<$ a $<$ 1250\,km, 540\,km $<$ b $<$ 759\,km, and 430\,km $<$ c $<$ 498\,km
(ellipsoidal axes a $>$ b $>$ c) for a triaxial rotationally deformed Haumea
with a homogeneous geometric albedo of p$_V$ = 0.6 -- 0.7.
\citet{Stansberry2008} conducted Spitzer-MIPS 70-$\mu$m observations and
combined it with visible photometry. Their radiometric analysis resulted in a
geometric albedo of 0.84$^{+0.10}_{-0.20}$ and a Spitzer-derived radiometric size of
1150$_{-100}^{+250}$ km. They also list a size of 1350 $\pm$ 100\,km refering
to \citet{Rabinowitz2005}.
\citet{Lellouch2010} combined the Spitzer-MIPS data with Herschel-PACS
measurements at 100 and 160 $\mu$m. They found an equivalent diameter (D$_{equ}$)
of $\sim$1300\,km (radiometric diameters between 1230 and 1324\,km)
and a geometric albedo of 0.70-0.81. The 100-$\mu$m thermal lightcurve
was explained by an elongated body with an axis ratio a/b =$\sim$1.3 and
a low thermal inertia of 0.2-0.5\,J~m$^{-2}$ s$^{-1/2}$ K$^{-1}$
(NEATM $\eta$ $<$ 1.15-1.35; for more information on the NEATM and the beaming
parameter $\eta$ see \citealt{Harris1998}).
\citet{Fornasier2013} included additional Herschel measurements at 70, 250, and
350 $\mu$m. They derived a radiometric size of 1239.5$^{+68.7}_{-57.8}$ km and a
geometric albedo $p_V$ of 0.804$^{+0.062}_{-0.095}$ based on a NEATM fit with
$\eta =0.95^{+0.33}_{-0.26}$. 

\citet{Lockwood2014} combined system-resolved HST data with re-analyzed
Spitzer-MIPS 70-$\mu$m data. They concluded that Haumea's extreme shape is the
cause of the large amplitude optical lightcurve, but also some longitudinal
surface heterogeneity is needed to explain the different lightcurve minima and 
maxima. The axis of their best-fit Jacobi triaxial ellipsoid had lengths of
1,920 $\times$ 1,540 $\times$ 990\,km (2a $\times$ 2b $\times$ 2c) and a high
density of about 2.6 g cm$^{-3}$.  \citet{Santos-Sanz2017} presented 100- and
160-$\mu$m thermal lightcurves of
Haumea obtained by the Herschel-PACS instrument apparently also showing the
100 $\mu$m lightcurve asymmetry connected to the DRS. The radiometric studies
implied a low thermal inertia ($<$0.5\,J~m$^{-2}$ s$^{-1/2}$ K$^{-1}$)
and a phase integral\footnote{The phase integral q allows to calculate
an object's bolometric Bond albedo A: A = p$\cdot$q, with p being the 
bolometric geometric albedo, assumed to be equal to the V-band albedo p$_V$.}
larger than 0.73 for Haumea's surface.
The best size estimate is given for a triaxial ellipsoid with a = 961\,km,
b = 768\,km, c = 499\,km, leading to a "mean area-equivalent diameter"
of 1309\,km, and $p_V = 0.71$, in agreement with previous radiometric studies.
\citet{Vilenius2018} used the
occultation results together with previously published fluxes by \citet{Fornasier2013}
to determine a NEATM $\eta$ of 1.74. Combining $\eta$ with the thermal parameter
$\Theta$ via the $\eta$-$\Theta$-relation \citep{Lellouch2013} gave a thermal inertia
estimate of $\Gamma$ $\sim$ 1\,J m$^{-2}$ s$^{-1/2}$ K$^{-1}$. Satellite and ring
contributions were neglected in previous aforementioned studies.

The recent occultation results \citep{Ortiz2017} are in
contradiction to the published radiometric properties (size, albedo, thermal inertia)
and the hydrostatic equilibrium
assumption. The occultation event combined with the optical lightcurve amplitude
leads to a much larger body size with about 2,322\,km for Haumea's longest axis (2a),
implying a density below 1.9\,g~cm$^{-3}$ (for the derived equivalent diameter of
1595\,km) and a geometric albedo of only 0.51.

In this work, we present new and updated infrared measurements and reinterpret the object's
thermal infrared emission in the light of the occultation results. 
In section \ref{sec:obs} we present the published thermal data set from Spitzer and
Herschel-SPIRE, and also the re-analyzed data from Herschel-PACS. In section \ref{sec:tpm}
we look at Haumea's spin-shape solution \citep{Ortiz2017}, derived from the multi-chord
occultation, and re-interpret the corresponding thermal measurements.
In section \ref{sec:discussion} we discuss our results and put them in context with
other icy bodies in the Solar System. Conclusions and an outlook to future studies
will be given in section \ref{sec:conclusion}.

\section{Thermal infrared observations}
\label{sec:obs}


\subsection{Spitzer-MIPS observations in 2005/2007}

Spitzer-MIPS \citep{Werner2004,Rieke2004} 24/70-$\mu$m measurements 
were taken on 2005 June 20/22 (listed in the Spitzer Heritage Archive under
"Santa", AORKEYs 13803008, 13802752) and 2007 July 13-19 (AORKEYs 19179520,
19179776, 19180032). \citet{Stansberry2008} published an upper limit of 0.022 mJy
in the 24-$\mu$m band and 7.8 mJy (S/N of 5.3) in the 70-$\mu$m band, derived
from the two 25-min measurement in 2005. A new data reduction
confirmed the 24-$\mu$m non-detection (in 13803008) and gave a marginal (S/N$\sim$2)
24-$\mu$m detection at 0.026 $\pm$ 0.012 mJy (individual background-subtracted
mosaics for 13802752). At 70 $\mu$m, the new calibrated in-band fluxes are
7.6 $\pm$ 1.6 mJy (from background-subtracted combined images) or
10.9 $\pm$ 1.8 mJy (13802752) and 4.3$\pm$2.1 mJy (13803008) for the individual
background-subtracted mosaics. The two measurements are separated by about
65 degrees in rotational phase, which makes the separate fluxes more useful.

In the case of the three 176-min observations in 2007, a recent reduction by
\citet{Lockwood2014} combined the data in blocks of 44 min of the same
rotational phases by using the 3.915341$\pm$0.000005 h rotation period from
\citet{Lellouch2010}. The new 70-$\mu$m fluxes are listed in Table~2 in
\citet{Lockwood2014}, but times are given in the Haumea reference frame. We
added 423.45\,min light-travel time to obtain the Spitzer-centric measurement
times (for compatibility with other observations, our IR database
requirements, and our model setup).
The observed MIPS thermal flux increase and decrease were found to be
positively correlated (on a 97\% level) with the HST optical lightcurve,
indicating a shape-driven origin. The \citet{Lockwood2014} 70-$\mu$m fluxes
show a fitted minimum level close to 12\,mJy and an amplitude of 4\,mJy.
However, the fit was done via a Jacobi ellipsoid
(1920$\times$1540$\times$990\,km) and several of the MIPS points deviate
considerably, with the lowest values around 8\,mJy and the highest
close to 20\,mJy.

The color correction for Haumea-like SEDs was estimated to be 1.28 for the
24-$\mu$m band (see also Table~4.16 in the MIPS Instrument
Handbook\footnote{\url{http://irsa.ipac.caltech.edu/data/SPITZER/docs/mips/mipsinstrumenthandbook/}}
for blackbodies in the temperature range between 30 and 50\,K). The color correction
at 70\,$\mu$m is 0.89 (see also \citealt{Stansberry2007}, Table~2, color
corrections for black bodies with temperatures in the range between 30 and 60\,K).
We divided the above fluxes by these factors to obtain mono-chromatic flux
densities at the 23.68 and 71.42-$\mu$m reference wavelengths. In addition, we
added 6\% calibration uncertainty to all MIPS flux errors when combining Spitzer
with Herschel measurements for radiometric size and albedo determinations, as 
recommended in \citet{Lockwood2014}.

In \citet{Fornasier2013} the lightcurve-averaged and color-corrected flux density
at 71.42\,$\mu$m is given with 15.83 $\pm$ 1.20\,mJy, in good agreement with
the averaged \citet{Lockwood2014} fluxes when taking the color-correction factor
of 0.89 into account.

\subsection{Herschel-PACS/-SPIRE observations of Haumea}

The Herschel Space Observatory \citep{Pilbratt2010} also performed a range of
measurements on Haumea with the PACS \citep{Poglitsch2010} and SPIRE
\citep{Griffin2010} instruments in the framework of the large key project
"TNOs-are-Cool!" \citep{Mueller2009}.
Subset of these data were presented in \citet{Lellouch2010}, \citet{Fornasier2013}, and \citet{Santos-Sanz2017}. 

\paragraph{Herschel-SPIRE}

For our analysis we took the SPIRE measurements from 2011 January 7/9 as
presented by \citet{Fornasier2013}. The two 35-min observation blocks were
combined to eliminate the strong background structures. The measurements
are separated by 49.94 hour which corresponds to almost exactly 270 degrees
in rotational phase, but both epochs are close to the mid-flux in the optical
lightcurve and can therefore be combined.

\paragraph{Herschel-PACS}

The Herschel-PACS observation were taken on 2009 December 23/25 (OBSIDs 1342188470, 1342188520; 100/160 $\mu$m),
and 2010 June 20/21 (1342198851, 1342198903/904/905/906; 70/100/160 $\mu$m).
\citet{Lellouch2010} presented an analysis of the 2009 lightcurve data, and
\citet{Fornasier2013} of the 2010 data. \citet{Santos-Sanz2017} focused on the
two lightcurve data sets from 2009 and 2010.

In the present paper, we re-analyzed and
re-calibrated all PACS observations with the latest tools and techniques
(\citealt{Kiss2014}).
The reprocessing includes a better noise characterization, the elimination of
the background which is very critical for faint sources in the two long-wavelength
channels at 100 and 160\,$\mu$m, an object-centered stacking of multiple measurements,
and a better understanding of the flux calibration for faint sources
(see also \citealt{Klaas2018}; \citealt{Balog2014}). The flux errors were determined
by performing aperture photometry at random places in the close vicinity of Haumea
and then combining with the PACS photometer intrinsic absolute flux calibration
error of 5\% \citep{Balog2014}. The final Haumea maps - together with all
"TNOs-are-Cool" observations (more than 400\,h of Herschel observing time; \citealt{Mueller2009}) -
were uploaded to the HSA\footnote{Herschel Science Archive} as "User Provided Data Products
(UPDP\footnote{\url{https://www.cosmos.esa.int/web/herschel/user-provided-data-products}})".
The details of the procedure and the new products are given in two specific release notes:
"User Provided Data Products of Herschel/PACS photometric light
curve measurements of trans-Neptunian objects and
Centaurs\footnote{\url{http://www.mpe.mpg.de/~tmueller/sbnaf/doc/tnosarecool\_hsa\_upload\_v1.pdf}}",
and "User Provided Data Products from the ”TNOs are Cool!  – A Survey of the trans-Neptunian
Region” Herschel Open Time Key Program\footnote{\url{http://www.mpe.mpg.de/~tmueller/sbnaf/doc/tno\_lightcurve\_hsa\_upload\_v1.pdf}}".

\begin{table*}
{\small
\caption{Epochs and observing geometry for all thermal measurements. PACS and SPIRE are Herschel instruments,
the observations are labeled with the Observation ID (OBSID), and the photometric bands are at 70, 100, 160,
250, 350, and 500\,$\mu$m. MIPS is a Spitzer instrument, the observations are connected to an AORKEY, the
relevant bands are at 24 and 70\,$\mu$m. The Haumea heliocentric distance (r$_{helio}$), the observatory-Haumea
distance ($\Delta$), and the phase angle $\alpha$ are given for the observation mid-time. The one-way light-travel
times vary between 421.97 and 426.33\,minutes. Note, that for the flux determination it was necessary to combine
different measurements (see text), the corresponding observing epochs in Tables~\ref{app:lc100a}, \ref{app:lc100b}, and
\ref{app:lc160} are therefore different. Observations were taken from the Spitzer Heritage Archive (SHA) and the
Herschel Science Archive (HSA). The PACS data are based on 10 individual 3.5(or 3.0)\,arcmin satellite scans,
separated by 10\,arcsec and either in 70 or 110$^{\circ}$ scan angles with respect to the instrument reference frame.
the lightcurve background measurement. The 1-3\,days separation of the measurements allows to reconstruct
mutual backgrounds while Haumea is always in the center of the map.}
    \label{tbl:obs}
\begin{tabular}{lrrcrccccl}
\hline\noalign{\smallskip}
            &            &         & Start     & Duration & mid-time JD  & r$_{helio}$ & $\Delta$ & $\alpha$    & \\
Instrument  & Identifier & Band(s) & time UT   & [sec]    & (+2450000.0) & [AU]        & [AU]     & [$^{\circ}$] & Comments \\
\noalign{\smallskip}\hline\noalign{\smallskip}                                          
MIPS   & 13803008   & 24/70       & 2005-Jun-20 17:20:37 &  1477 & 3542.23125 & 51.242 & 50.906 & 1.07 & "Santa" in SHA     \\ 
MIPS   & 13802752   & 24/70       & 2005-Jun-22 09:11:29 &  1477 & 3543.89167 & 51.242 & 50.929 & 1.08 & "Santa" in SHA     \\ 
MIPS   & 19179520   & 24/70       & 2007-Jul-13 11:01:22 & 10294 & 4295.01886 & 51.152 & 50.885 & 1.10 & "2003 EL61" in SHA \\ 
MIPS   & 19179776   & 24/70       & 2007-Jul-16 04:23:05 & 10297 & 4297.74229 & 51.152 & 50.924 & 1.11 & "2003 EL61" in SHA \\ 
MIPS   & 19180032   & 24/70       & 2007-Jul-19 06:43:33 & 10296 & 4300.83983 & 51.152 & 50.969 & 1.12 & "2003 EL61" in SHA \\ 
PACS  & 1342188470 & 100/160     & 2009-Dec-23 05:52:01 & 12084 & 5188.81439 & 51.028 & 51.262 & 1.08 & lightcurve 1       \\ 
PACS  & 1342188520 & 100/160     & 2009-Dec-25 06:13:39 &  2420 & 5190.77348 & 51.028 & 51.232 & 1.09 & background         \\ 
PACS  & 1342198851 & 100/160     & 2010-Jun-20 20:45:11 & 15514 & 5368.45449 & 51.001 & 50.737 & 1.12 & lightcurve 2       \\ 
PACS  & 1342198903 &  70/160     & 2010-Jun-21 22:42:00 &   568 & 5369.44912 & 51.001 & 50.752 & 1.12 & scan map           \\ 
PACS  & 1342198904 &  70/160     & 2010-Jun-21 22:52:31 &   568 & 5369.45642 & 51.001 & 50.752 & 1.12 & cross-scan map     \\ 
PACS  & 1342198905 & 100/160     & 2010-Jun-21 23:03:02 &   568 & 5369.46373 & 51.001 & 50.752 & 1.12 & scan map           \\ 
PACS  & 1342198906 & 100/160     & 2010-Jun-21 23:13:33 &   568 & 5369.47103 & 51.001 & 50.752 & 1.12 & cross-scan map     \\ 
SPIRE & 1342212360 & 250/350/500 & 2011-Jan-07 07:08:50 &  2111 & 5568.81002 & 50.971 & 50.994 & 1.12 & small scan map     \\ 
SPIRE & 1342212414 & 250/350/500 & 2011-Jan-09 09:05:18 &  2111 & 5570.89090 & 50.971 & 50.962 & 1.12 & follow-on map      \\ 
\noalign{\smallskip}\hline\noalign{\smallskip}
\end{tabular}
}
\end{table*}

Two sets of 100/160-$\mu$m lightcurves were taken: (1) A 3h21min sequence on
2009 December 23, followed by a $\sim$40\,min background measurement on 2009 Dec 25. 
However, Haumea had moved by more than 80\,arcsec between both measurements (Haumea's
apparent motion was about 1.7 arcsec/hour) which means that the relevant background
region had moved to the edge of the 105$\times$210\,arcsec FOV of the PACS bolometer
array in the follow-up measurements on Dec 25. Haumea's sky background in the long-duration
measurement is therefore only poorly characterized and part of the
observed flux variations are very likely related to the structured background
along Haumea's path. 
(2) A 4h18min sequence on 2010 June 20. In this second sequence, Haumea moved
about 5.6 arcsec on the sky during the 4.3 hours (compared to the $\sim$7/12 arcsec
FWHM of a point-source PSF at 100/160 $\mu$m, respectively). The sky background
in these two bands were taken from the measurements taken on the following day, when
Haumea had moved about 30 arcsec. We performed aperture photometry on the background-subtracted
images, and color-corrected the resulting fluxes (see also \citealt{Santos-Sanz2017}).
For the 100 and 160\,$\mu$m lightcurve data we merged 3 and 6 repetitions, respectively.
For the overlap, the shift between the 100\,$\mu$m data points is 1 repetition, but it
is 3 repetitions between the 160\,$\mu$m data points, leading to an oversampled
of a factor of 3(6) at 100(160)\,$\mu$m.
The final numbers are given in Tables~\ref{app:lc100a}, \ref{app:lc100b}, and
\ref{app:lc160}.

\begin{figure*}
  \centering
  \includegraphics[angle=0,width=0.70\hsize]{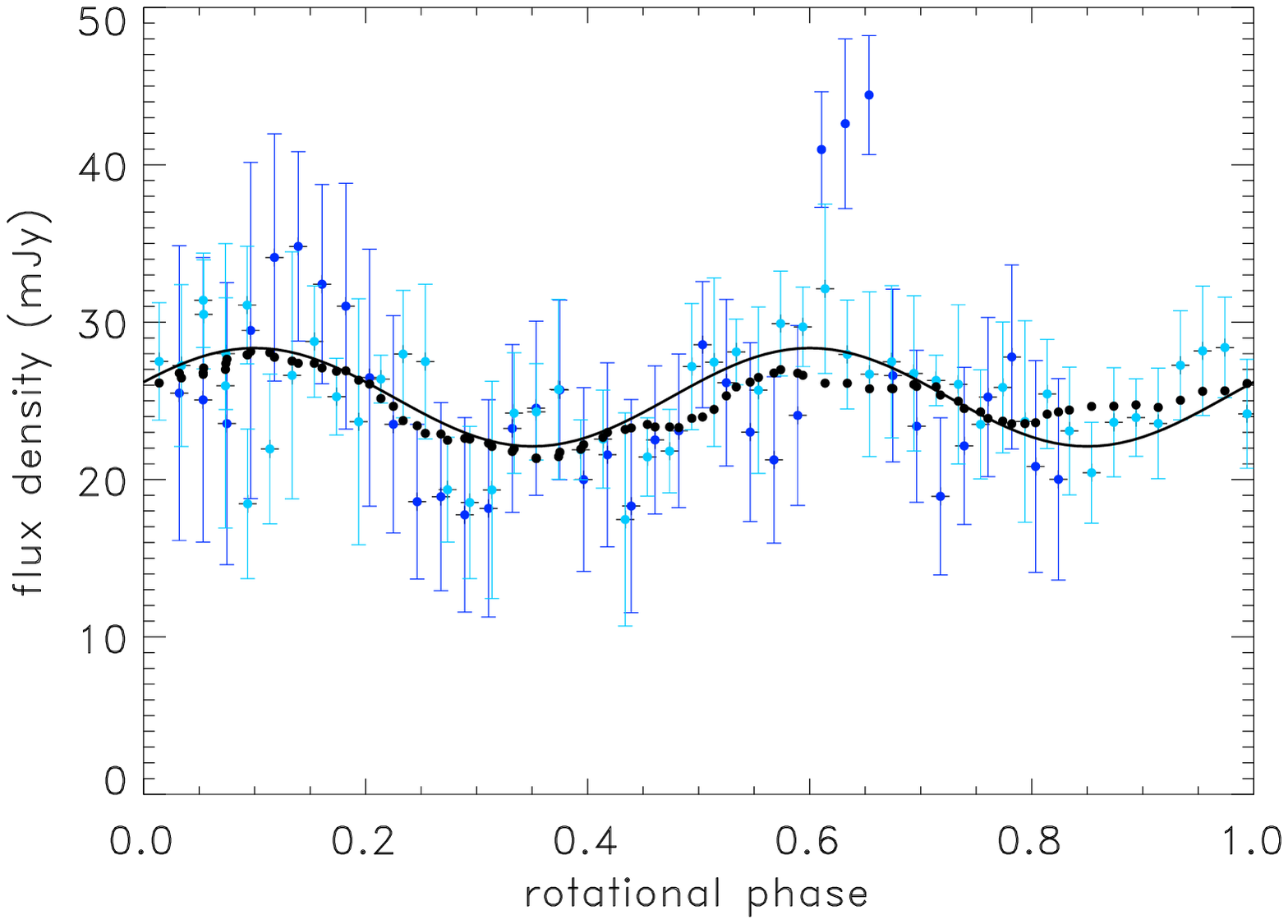}
  \includegraphics[angle=0,width=0.70\hsize]{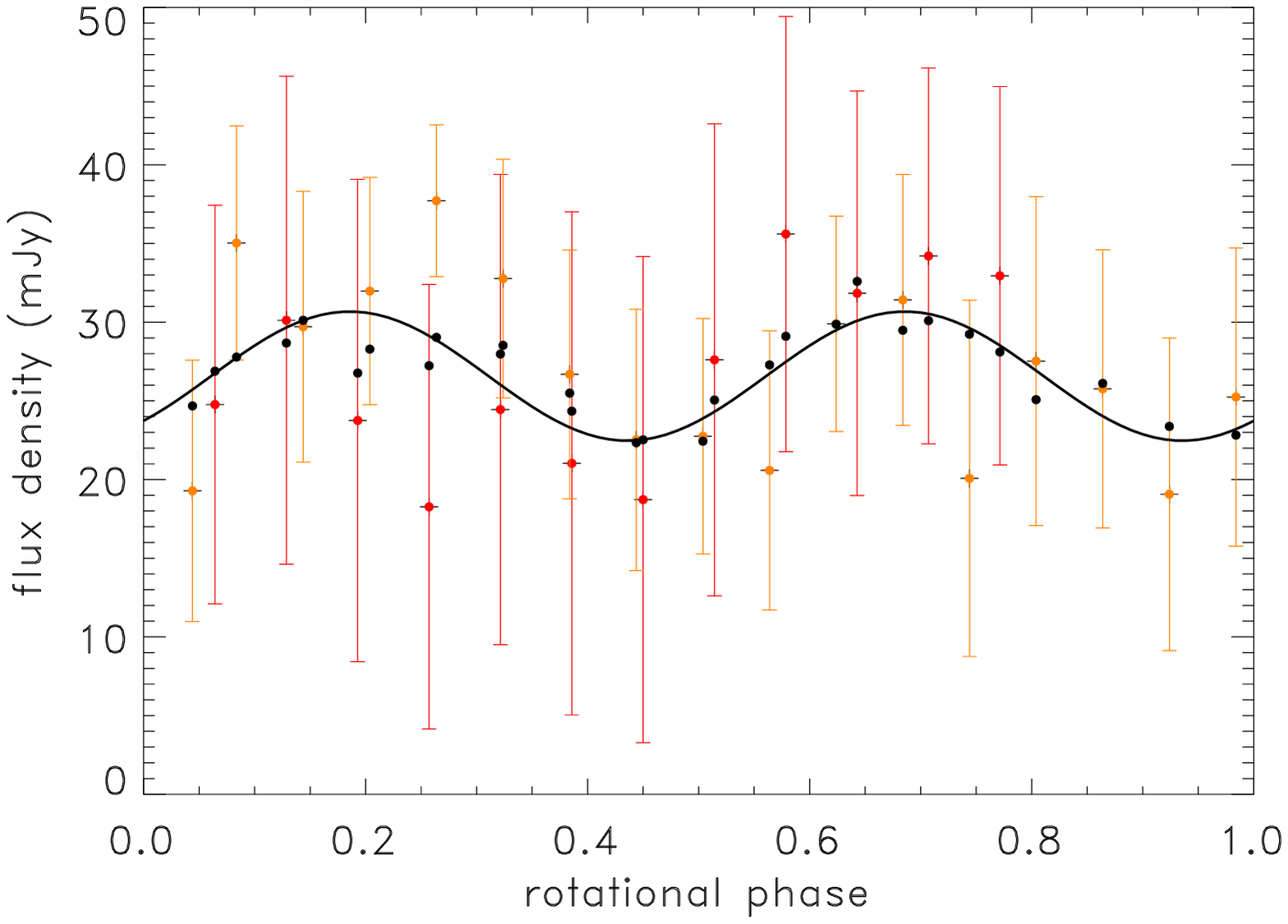}
  \caption{The two sets of 100- (top) and 160-$\mu$m (bottom) Herschel-PACS lightcurve
  measurements combined (see text for details). The zero phase corresponds to JD = 2455188.743985.
  The dark blue(red) points are of the 1$^{st}$ epoch, and the light blue(orange) ones are
  of the 2$^{nd}$ epoch. The solid black curve is the sinusoidal fit. The curve
  drawn by the black dots is a smooth curve from the original data (the fit is
  made with the original data, not the smoothed one, without
  the three outliers around $\phi$=0.65). The 100\,$\mu$m lightcurve seems to
  be asymmetric with the lower peak at $\phi$=0.6.}\label{fig:PacsLcData}
\end{figure*}

With the knowledge of Haumea's precise rotation period, it is possible to
combine the 100/160-$\mu$m lightcurves from December 2009 and June 2010 to 
plot the obtained fluxes as a function of rotational phase using P$_{rot}$ =
3.915341\,h \citep{Santos-Sanz2017, Thirouin2013PhD}, and a (Herschel-centric) reference date of
JD = 2455188.743985 (phase $\phi$=0), the start time of the observations.
It is worth to note that the light travel times and the rotational phase shifts
between the two lightcurve epochs are only 13.5\,s and 0.0009, respectively,
much shorter than typical measurement time of a few minutes.
Figure~\ref{fig:PacsLcData}
shows the phase-folded 100- and 160-$\mu$m data, respectively.

The 100-$\mu$m double-peaked sinusoidal fit gave A$_0$ = 25.23$\pm$0.89 mJy and A$_1$
= 3.12$\pm$1.25 mJy (Flux = A$_{0}$ + A$_{1}$\,sin($\phi$)).
Overall, the consideration of the 1$^{st}$ lightcurve from
December 2009 seems to increase the lightcurve amplitude a bit, but it gives
almost exactly the same mean (A$_0$).
In the caption of Fig.~\ref{fig:PacsLcData} we explain the data and model fits.
We see a definite lightcurve at 100 $\mu$m, and it is likely asymmetric (lower
peak at $\phi$=0.6). There are some data points from the 1$^{st}$ lightcurve data (dark blue)
that seemingly increase the amplitude, but these points are affected by single
repetitions that provide a very high flux density. They all happen in the 1$^{st}$
lightcurve with the problematic background.

The combined light curve at 160 $\mu$m shows the first-epoch data in red and
the second-epoch data in orange. The error bars are much larger at the 1$^{st}$
epoch due to the background characterization problems. We determined 
A$_0$=26.57$\pm$2.73\,mJy and A$_1$=4.10$\pm$3.85\,mJy. The uncertainty on A$_1$ is very
large, but produced by our $\chi^2$ analysis of the data. We also tried to refit the 160\,$\mu$m
data with a fixed lightcurve phase, as obtained from the 100\,$\mu$m lightcurve. In this case,
the mean is the same, and the lightcurve amplitude is $<$2.5\,mJy, with a nearly
flat (uniform) distribution between 0 and 2.5\,mJy (formally the highest
probability is at $\approx$1.63\,mJy).
It is worth to note that both, our 100\,$\mu$m and 160\,$\mu$m lightcurve amplitudes,
correspond to $\sim$0.3\,mag, very close to the visible lightcurve amplitude.

The four short 10-min PACS measurements from 2010 June 21 (see Table~\ref{tbl:obs})
are reduced in a standard way by combining the scan- and cross-scan images.
The 100- and 160-$\mu$m background maps (from 2010 June 20) were subtracted.
The 70-$\mu$m map has no dedicated background
counterpart, but background confusion is less severe at these shorter
wavelengths.
The extracted fluxes are aperture and color corrected and come with a proper noise determination.
The fluxes are given in the tables in the appendix (Tables~\ref{app:lc100a}, \ref{app:lc100b}, \ref{app:lc160}, \ref{app:phot}).
These 3-band data can also be phased into our lightcurve data: The corresponding rotation
phase is around 0.75, which is very close to the mid-flux level.

Overall, the new reduction procedure increased the reliability of the extracted
fluxes and reduced the previously published peak-to-peak lightcurve amplitudes
significantly. The poorly characterized background structures in the 2009
lightcurve data were the main cause for the large flux variation shown in
\citet{Lellouch2010} (peak-to-peak lightcurve amplitude of $\sim$17\,mJy
at 100\,$\mu$m) and \citet{Santos-Sanz2017} (about 10(8)\,mJy variation
at 100(160)\,$\mu$m).

\section{Thermophysical model calculations}
\label{sec:tpm}

The radiometric technique (see e.g.\ \citealt{Delbo2015} and references therein)
usually takes measured thermal infrared fluxes to derive size, albedo, and
thermal properties for a given object. In the case of Haumea, we follow a different
approach, mainly because of the direct size measurement from the occultation
event, and the unknowns in the satellite-ring contributions to the thermal signals.

\subsection{Using the occultation cross-section}
\label{sec:tpmocc1}
We use the occultation-derived 2-D ellipse fit for Haumea, as measured on 2017
January 21 at around 03:09 UT ($\pm$ about 1.5 min). Haumea's projected shape
was fit by an ellipse with a=852$\pm$2 km and b=569$\pm$13 km or a
circle-equivalent diameter of $2\sqrt{852\times 569} =$1392.5 km ($\pm$10.2 km)
\citep{Ortiz2017}.
The lightcurves taken close in time to the occultation event showed that Haumea
was very close to its minimum brightness, both in reflected light and thermal
emission (see \citealt{Lockwood2014}; \citealt{Santos-Sanz2017}).
Regarding the DRS, this is centered in one of the lightcurve maxima
\citep{Lacerda2008},
which means that at lightcurve minimum the DRS would be in grazing view
from the Earth with a tiny effect on the thermal flux.

In a first step, we use only the occultation cross section and the object's rotational
properties \citep{Lellouch2010,Ortiz2017} to make thermophysical predictions
at the Spitzer and Herschel observing epochs and geometries. The following
(relevant) parameters were used in our thermophysical model (TPM) setup:

\begin{itemize}
\item For the interpretation of the fitted occultation ellipse we use an
      oblate spheroid with a = b = 852\,km; c = 513\,km which resembles the
      occultation ellipse when projecting the oblate spheroid under an
      aspect angle of 76.2$^{\circ}$.
\item We use the occultation-derived rotation poles at
      (RA, Dec)=(285.1$^{\circ}$, -10.6$^{\circ}$) and
      (RA, Dec)=(312.3$^{\circ}$, -18.6$^{\circ}$),
      named pole 1 and pole 2, respectively (in ecliptic coordinates:
      (285.2$^{\circ}$, +12.1$^{\circ}$) and (309.6$^{\circ}$, -0.8$^{\circ}$));
      however, pole 1 was given priority in \citet{Ortiz2017} because of the long-term 
      lightcurve amplitude behavior of Haumea and the agreement with Hi'iaka's orbital pole position.
\item The object's rotation period was found to be P$_{rot}$ = 3.915341 hours \citep{Santos-Sanz2017}.
\item The rotationally averaged absolute magnitude of the Haumea-system is $H_V=$+0.35$\pm$0.06,
      but we use the $H_V$ value of Haumea's main body at the time of the occultation:
      0.35 + 0.14 + 0.32/2\,mag  (to account for the satellites and the lightcurve
      amplitude; \citealt{Ortiz2017}) = 0.65 mag.
\item Then, the object's geometric albedo can be calculated \citep{Harris1998}:
      \begin{equation}
        p_V = (1329/D(km))^2 \cdot 10^{(-0.4 \cdot H)}
      \end{equation}
      The resulting geometric albedo connected to the occultation ellipse size is then 0.51$\pm$0.02
      \citep{Ortiz2017} or $\pm$0.05 if we take the 10.2-km size error and a conservative $\pm$0.1\,mag error for H$_{mag}$
      into account.
\item For the phase integral we applied the formula by \citet{Brucker2009}:
      $q = 0.336 p_V + 0.479$. This led to $q = 0.65$ (for $p_V=0.51$).
      Whenever relevant, we tested also a wider range of $q= 0.45$ to 0.85 to account
      for the unknown scattering properties on Haumea's surface.
\item Like in most radiometric studies, we use a bolometric emissivity $\epsilon = 0.9$ in the MIPS/PACS range,
      and a lower emissivity of 0.8 at sub-millimeter wavelengths in the SPIRE range (see \citealt{Mueller2002},
      \citealt{Fornasier2013}; \citealt{Lellouch2016}; \citealt{Lellouch2017}).
\item We consider a very wide range of surface roughness values
      (as a free parameter) ranging between 0.1 (very smooth)
      and 0.9 (extremely rough)
      in r.m.s.\ of surface slopes. However, the lower range with r.m.s\ $<$0.5
      seems to be more realistic for TNOs (see findings in \citet{Fornasier2013}).
\item The thermal inertia is our free parameter and we test a range between
      0.01\,J K$^{-1}$ m$^{-2}$s$^{-1/2}$ (extremely low thermal conductivity) and
      100\,J K$^{-1}$ m$^{-2}$s$^{-1/2}$ (extremely high conductivity considering the
      low-temperature environment at 50\,AU).
\item The differences in Spitzer-Haumea and Herschel-Haumea distances and
      phase angles are small and lead to negligible differences in the
      TPM fluxes of only 1-2\% (assuming low thermal inertia below 5\,J K$^{-1}$ m$^{-2}$s$^{-1/2}$)
      or less for thermal inertias $>$5\,J K$^{-1}$ m$^{-2}$s$^{-1/2}$, for the 6-yr span of 
      thermal observations (Tbl.~ref{tbl:obs}).
\end{itemize}

\begin{figure*}
  \centering
  \includegraphics[angle=90,width=0.70\hsize]{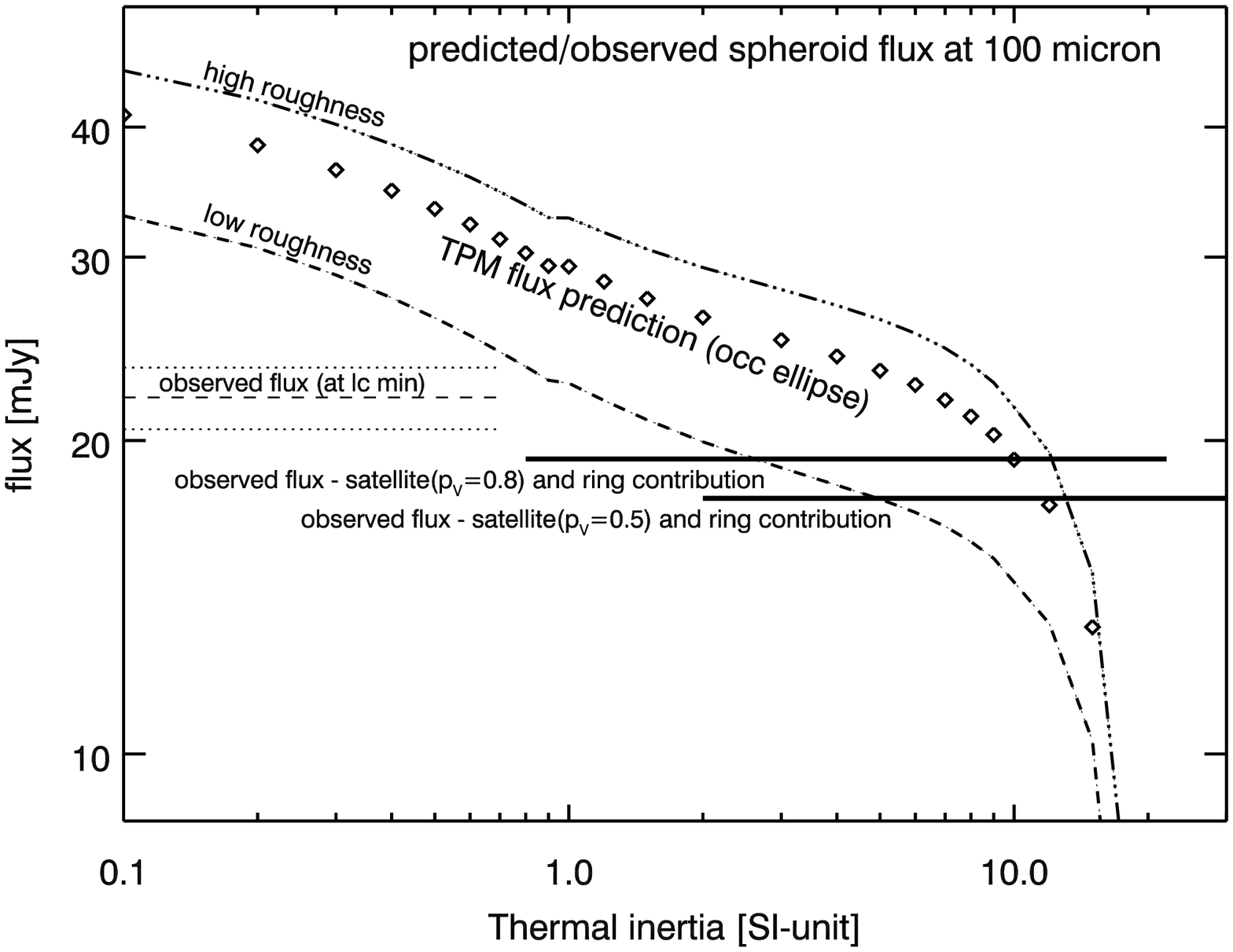}
  \includegraphics[angle=90,width=0.70\hsize]{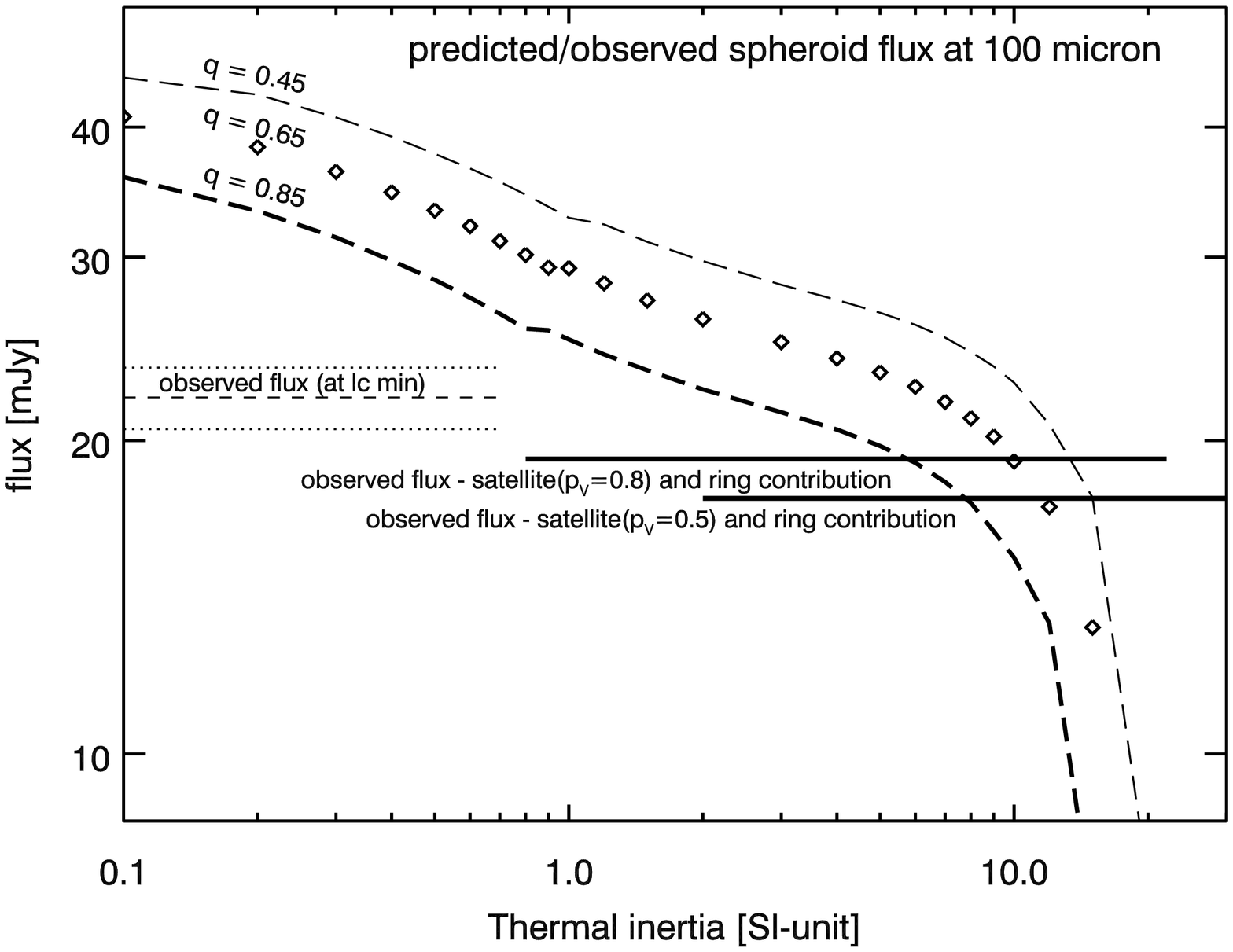}
  \caption{TPM 100-$\mu$m flux calculations using only the occultation 2-D ellipse information.
           The measured 100-$\mu$m flux (of the thermal lightcurve minimum) is over-plotted:
           Left side within the figure: the observed Haumea-ring-satellite fluxes (with errors
           shown as dotted lines); right side: the estimated
           satellite-ring contribution was subtracted from the observed flux, assuming an albedo of p$_V$=0.8 for
           the satellites (horizontal top solid/dashed lines), and p$_V$=0.5 (lower solid line).
           Top: Using a wide range of extreme roughness levels on the surface.
           Bottom: Using a wide range of phase integrals (from 0.45 to 0.85).
           The intersections between model and observed fluxes indicate that
           Haumea's thermal inertia must be above
           $\sim$2 and below $\sim$20\,J m$^{-2}$ s$^{-1/2}$ K$^{-1}$}.
           \label{fig:TPM100um1}
\end{figure*}

Using the above settings, we can make flux predictions for the full range
of thermal inertias and roughness levels for the specific occultation
size. The calculations are done via a TPM \citep{Lagerros1996I,
Lagerros1997, Lagerros1998, Mueller1998, Mueller2002} and, in the first step, by using a
spheroid resembling the fitted occultation ellipse (see explanation above).
The TPM predictions are shown in Fig.~\ref{fig:TPM100um1}.
The illuminated side of a TNO coincides with the visible hemisphere given
the typical small phase angle of the observations. However,
the thermal fluxes (default values are shown as diamonds) decrease with
increasing thermal inertias (more heat is transported to the night side).
At $\sim$10\,J K$^{-1}$ m$^{-2}$s$^{-1/2}$ Haumea is almost isothermal
(except at the poorly illuminated polar regions) with the 30-40\,K zone reaching
latitudes of above 60$^{\circ}$.
At much larger thermal inertias more and more heat goes to the subsurface (lowering
the effective temperature) and the warm terrain shrinks towards the equatorial zone
(still isothermal at each latitude) which explains the steep decrease at very
high thermal inertias.
The assumptions of a high/low surface roughness (dashed line below and dashed-dotted
line above the diamonds) produces higher/lower fluxes, but the influence of roughness
shrinks with increasing thermal inertia.
We also tested the influence of the phase integral q (0.45$<$q$<$0.85), but the lines are
within the low/high roughness corridor, with the q=0.45 prediction close to the
"high roughness" levels, and the q=0.85 predictions close to the "low roughness"
dashed line (see Fig.~\ref{fig:TPM100um1}, bottom).

These predictions can now be compared with the calibrated, color-corrected absolute
flux densities.
We take the fluxes at thermal lightcurve minimum since the occultation happened
close to the minimum \citep{Santos-Sanz2017}: $\approx$12\,mJy at 70\,$\mu$m
(estimated from Figure~1 in \citet{Lockwood2014}), 22$\pm$1.5\,mJy at 100\,$\mu$m,
(horizontal dashed and dotted lines on the left side in Figure~\ref{fig:TPM100um1}
top and bottom part),
and 22.5$\pm$3\,mJy at 160\,$\mu$m (see fit parameters A$_0$ and A$_1$ above).
But there is one problem with that comparison: the thermal measurements
include contributions from the ring and Haumea's satellites
(see discussion in Section~\ref{sec:satsring}).
The Haumea-only fluxes (subtracting the smallest possible ring-satellite
contribution) are also shown in the figure as thicker solid
lines on the right side of Figure~\ref{fig:TPM100um1}. The intersections with the flux curves indicate the
most probable thermal inertia for Haumea itself. These simple calculations are
based on the measured occultation ellipse and Haumea's H-magnitude and rotational
properties, but they show that the occultation result is only compatible with
the thermal measurements when the thermal inertia is larger than
$\approx$2\,J\,K$^{-1}$\,m$^{-2}$\,s$^{-1/2}$. This lower limit for the
thermal inertia is confirmed by similar exercises at 70, 160, and 250\,$\mu$m
where we also have multiple observations. The figure shows that
higher ring/satellite contributions (e.g., when assuming a less-extreme lower
albedo for the satellites) would automatically lead to even higher thermal
inertias for Haumea. Also the TPM assumptions for the surface roughness 
play a role: for a low-roughness (smooth) surface we estimate a thermal
inertia of 2 (or higher), while a very rough surface would give also
higher thermal inertias well above 10\,J\,K$^{-1}$\,m$^{-2}$\,s$^{-1/2}$.
TPM analysis of other targets (\citealt{Fornasier2013}; \citealt{Kiss2018})
indicated that lower roughness values (r.m.s.\ of surface slopes between
0.1 to 0.3) seem to fit better the thermal signals of dwarf planets.
Looking at the influence of the phase integral shows an opposite trend:
high q-values are more compatible with low thermal inertias
(above $\sim$5\,J\,K$^{-1}$\,m$^{-2}$\,s$^{-1/2}$), while q-values below
0.5 would require Haumea to have $\Gamma$ well above 10\,J\,K$^{-1}$\,m$^{-2}$\,s$^{-1/2}$
(Figure~\ref{fig:TPM100um1} bottom part).

\subsection{Using the occultation-lightcurve derived 3-D size-spin-shape solution}
\label{sec:tpmocc2}

\begin{figure*}
  \centering
  \includegraphics[angle=90,width=0.73\hsize]{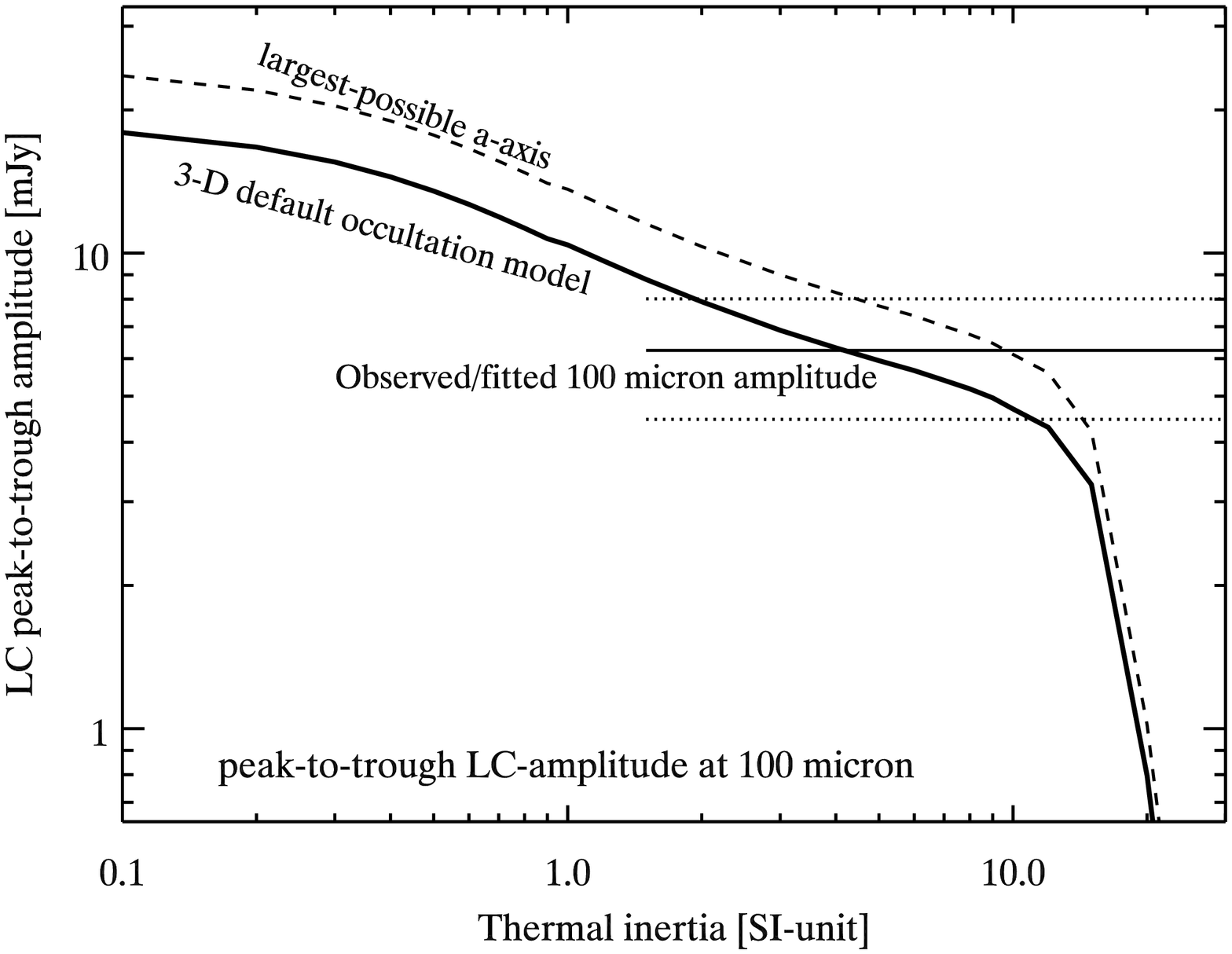}
  \caption{TPM 100-$\mu$m lightcurve amplitudes for the default 3-D spin-shape-size
           solution derived from the occultation (solid line) and using the largest possible
           a-axis extension (see \citealt{Ortiz2017}) (dashed line). The observed lightcurve
           amplitude is over-plotted (see Fig.~\ref{fig:PacsLcData}).
           This approach leads to a thermal inertia above $\sim$2 and below
           $\sim$15\,J m$^{-2}$ s$^{-1/2}$ K$^{-1}$ for Haumea.}
           \label{fig:TPM100um2}
\end{figure*}

The above exercise (Fig.~\ref{fig:TPM100um1}) used minimum fluxes in each band for a direct comparison with
the occultation 2-D results. But the five thermal ligthcurves (single MIPS 70\,$\mu$m;
two-epoch PACS 100/160-$\mu$m lightcurves) contain additional information.
The absolute fluxes are "contaminated" by the unknown ring/satellite contributions,
but the lightcurve amplitude is directly connected to Haumea itself \citep{Lockwood2014}.
For the interpretation of the thermal lightcurves we use the \citet{Ortiz2017}
3-axes ellipsoid with a default size of a= (1161$\pm$30)\,km, b=(852$\pm$2)\,km,
and c=(513$\pm$16)\,km (a/b=1.36 and b/c=1.66) and a volume-equivalent diameter
of $2 \cdot (1161 \cdot 852 \cdot 513)^{1/3}$ = 1595\,km. In addition, we also consider
the more extreme solution (based on a maximum of 5\% for the visual brightness
contribution of the ring to the observed magnitudes) with 2a = 2520\,km
(a/b = 1.48; D$_{equ}$ = 1632\,km), also given in \citet{Ortiz2017}.
The corresponding thermal lightcurve
amplitude decreases with increasing thermal inertia (see Figure~\ref{fig:TPM100um2}),
with only very little influence by the surface roughness assumptions or the value for
the phase integral (not shown in the figure).
The "default" occultation model is shown with a solid line, the extreme
case with 2a = 2520\,km produces a larger amplitude (shown as dashed line).
The 100-$\mu$m lightcurve amplitude which was fitted to the observed fluxes
(see Figure~\ref{fig:PacsLcData}) is over-plotted (solid horizontal line). The
formal amplitude error from the fit is also shown (dashed horizontal lines).
The 100\,$\mu$m lightcurve amplitude indicates that Haumea must have a 
thermal inertia in an estimated range between $\sim$2 and about 10\,J m$^{-2}$ s$^{-1/2}$ K$^{-1}$.
If the a-axis is more extended (dashed line in Fig.~\ref{fig:TPM100um2}) then
the limits are between 4 and about 15\,J m$^{-2}$ s$^{-1/2}$ K$^{-1}$.
The lightcurves at 70\,$\mu$m (MIPS) and 160\,$\mu$m give a very
similar picture, although the formal errors are bigger. Figure~\ref{fig:TPM100um2}
explains nicely why previous studies \citep{Lellouch2010,Santos-Sanz2017}
resulted in much smaller thermal inertias for Haumea. In these projects, the
light curve amplitudes were heavily overestimated due to the poorly characterised
background.
In addition, previous assumed shape models (based on hydrostatic equilibrium) were less
elongated than the one from the occultation, and the derived radiometric
size was smaller.
Both aspects pushed the derived thermal inertia to very small values.
Having the occultation-derived shape, spin pole and size, also allowed us to 
look into the shift between the optical and thermal lightcurves. We calculated
this shift for thermal inertias ranging from $\Gamma$=0\,J m$^{-2}$ s$^{-1/2}$ K$^{-1}$
(in phase with the optical lightcurve) to $\Gamma$=25\,J m$^{-2}$ s$^{-1/2}$ K$^{-1}$
(thermal lightcurve is delayed with respect to the optical). We determined delay times
(see also \citealt{Santos-Sanz2017}) of up to 5\,min (or $\sim$0.02 in rotational phase)
for the current aspect angle, but our observations are clearly not accurate
enough to measure this effect
directly. However, the optical-thermal delay might still be an interesting 
approach for other targets (with less-extreme shapes) to constrain the object's thermal inertia.

\subsection{Estimating the thermal contribution of the satellites and the ring}
\label{sec:satsring}

Haumea is known to have two satellites \citep{Brown2005,Brown2006} and a ring
\citep{Ortiz2017}. \citet{Ragozzine2009} determined the mass of
Haumea's satellites from the orbits in the framework of a three point mass
model (1.79$\pm$0.11 $\cdot$10$^{19}$\,kg for Hi'iaka, and
       1.79$\pm$1.48 $\cdot$10$^{18}$\,kg for Namaka).
Using a realistic density estimate of $\rho$\,=\,0.5--1.5\,g/cm$^3$
for both moons, the size ranges are D$_H$\,=\,350$\pm$50\,km, and
D$_N$\,=\,150$\pm$50\,km for Hi'iaka and Namaka, respectively.
Additionally, \citet{Brown2006} listed the moons' brightnesses to be
5.9+/-1.5\% and 1.5+/-0.5\% of Haumea's brightness, while \citet{Ragozzine2009}
reported Hi'iaka to be $\approx$10 times fainter than Haumea and Namaka $\approx$3.7
times fainter than Hi'iaka, corresponding to $\sim$3.5 and 5.1\,mag
for their absolute magnitudes\footnote{see also \url{http://www.johnstonsarchive.net/astro/astmoons/am-136108.html}, 
retrieved on Oct 24, 2018.}, given Haumea's absolute magnitude,
H$_V$ = 0.49\,mag \citep{Ortiz2017}.
We predicted the flux densities of Haumea's satellites using a NEATM
model \citep{Harris1998} for the above range of possible mass-related sizes
and albedos (fulfilling the H-magnitude constraint), and assuming the
\citet{Brucker2009} phase-integral-albedo relation.
We consider the
usage of the NEATM as appropriate in case of the satellites where no
spin properties (rotation period and/or spin-axis orientation) are known.
In Table~\ref{tbl:satflux} we list the corresponding flux predictions
assuming a beaming parameter of $\eta$\,=\,1. This $\eta$-value
is close to the weighted mean value of 1.07 derived from combined Herschel and Spitzer
observations of 85 TNOs and Centaurs \citep{Lellouch2013}.

\begin{table*}[!h]
  \centering
  \caption{NEATM predictions for Haumea's satellites, assuming a beaming parameter 
           $\eta$=1.0, and a constant emissivity of $\epsilon$=0.9.
           "PhaseInt": value for the phase integral.}\label{tbl:satflux}
  \begin{tabular}{l c c c c c c c c}
    \noalign{\smallskip}
    \hline
    \noalign{\smallskip}
    Diameter & Albedo & PhaseInt & \multicolumn{6}{c}{Hi'iaka flux (mJy) at} \\
    (km)     & p$_V$  & q       & 24$\mu$m & 70$\mu$m & 100$\mu$m & 160$\mu$m & 250$\mu$m & 350$\mu$m \\
    \noalign{\smallskip}
    \hline
    \noalign{\smallskip}
    300 & 0.79 & 0.74 &  0.002   & 0.86 & 1.33 & 1.31 & 0.89 & 0.58 \\ 
    350 & 0.58 & 0.67 &  0.011   & 1.85 & 2.52 & 2.24 & 1.44 & 0.91 \\ 
    400 & 0.44 & 0.63 &  0.023   & 2.93 & 3.78 & 3.22 & 2.03 & 1.27 \\ 
    \noalign{\smallskip}
    \hline
    \noalign{\smallskip}
    Diameter & Albedo & PInteg. & \multicolumn{6}{c}{Namaka flux (mJy) at} \\
    (km)     & p$_V$  & q       & 24$\mu$m & 70$\mu$m & 100$\mu$m & 160$\mu$m & 250$\mu$m & 350$\mu$m \\
    \noalign{\smallskip}
    \hline
    \noalign{\smallskip}
    140 & 0.83 & 0.76 & $<$0.001 & 0.17 & 0.26 & 0.27 & 0.18  & 0.12 \\ 
    160 & 0.64 & 0.69 & 0.002    & 0.35 & 0.49 & 0.44 & 0.29  & 0.18 \\ 
    180 & 0.50 & 0.65 & 0.004    & 0.55 & 0.72 & 0.63 & 0.40  & 0.25 \\ 
    200 & 0.41 & 0.62 & 0.006    & 0.76 & 0.97 & 0.82 & 0.51 & 0.32 \\ 
    \noalign{\smallskip}
    \hline
  \end{tabular}
\end{table*}

For both satellites, smaller sizes than the above stated limits would require
unrealistically high geometric albedos, p$_V$\,$>$\,1, which is not
expected for inactive bodies.
The two satellites give a minimum thermal emission contribution of
$\sim$1.0\,mJy at 70\,$\mu$m, 1.6\,mJy at 100\,$\mu$m, 1.6\,mJy at 160\,$\mu$m,
1.1\,mJy at 250\,$\mu$m, and 0.7\,mJy at 350\,$\mu$m (assuming p$_V$ $\sim$0.8
for both). For a lower albedo of p$_V$ $\sim$0.5, the contributions
at 70, 100, 160, 250, and 350\,$\mu$m are already 2.4, 3.2, 2.9, 1.8, and 1.2\,mJy,
respectively.
We also investigated the influence of a higher beaming paramter with $\eta$=1.3.
In this case, the satellites contribute about 20\% less flux in the PACS
and about 10\% less in the SPIRE range.

It is worth to note here that Haumea might have more
so-far undiscovered satellites, including potential shepherd satellites for the ring.
They might also add a small contribution to the observed fluxes of the
Haumea-ring-satellite system.

\begin{figure*}
  \centering
  \includegraphics[angle=270,width=0.73\hsize]{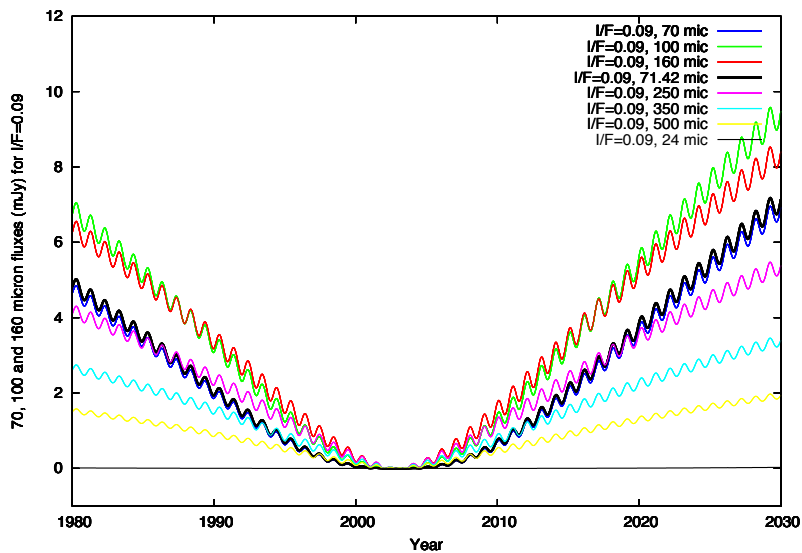}
  \caption{Thermal model flux predictions for Haumea's ring for the
           time period 1980 to 2030. The Spitzer observations happened
           in 2005 and 2007 when the ring contribution was negligible.
           For the interpretation of the Herschel observations (2009-2011)
           we took the 0.4 to 1.4\,mJy contribution into account, depending 
           on the filter and observing time.}\label{fig:ringpred}
\end{figure*}

To estimate the thermal contribution of Haumea's ring, we used the simple thermal model from
\citet{Lellouch2017}, itself based on a simplified version of a model for Saturn's rings.
In this model, the only source of energy for ring particles is absorbed solar radiation,
but mutual shadowing - as seen both from the Sun and the observer -
and optical depth effects are taken into consideration. The model further
assumes that ring particles have a bolometric and spectral emissivity of unity.
By analogy with Saturn's rings, the latter assumption is certainly valid
up to $\sim$200 $\mu$m, but the spectral emissivity could decrease at longer
wavelengths if ring particles are made of water ice. Model free parameters are the
ring radius $r$, width $w$, opacity $\tau$, and Bond albedo $A_b$, related to the
I/F reflectivity\footnote{I is the intensity of light reflected by the
surface and $\pi$F is the incident solar flux density.} 
and the phase integral, assumed here to be equal to 0.5. We use the nominal values
from \citep{Ortiz2017}, i.e.\ $r$ = 2287\,km, $w$ = 70\,km, $\tau$ = 0.5 and I/F = 0.09
(see also the discussion in \citealt{Braga-Ribas2014}).
Although the latter quantity is the least well constrained, it has a minor
effect on the ring thermal emission. We calculated the ring thermal flux for
all viewing geometries from 1980 to 2030 (see Figure~\ref{fig:ringpred}). 
The calculated thermal fluxes at some selected wavelengths and dates are
listed in Table~\ref{tbl:ringflux}.

\begin{table}
  \centering
  \caption{Thermal emission predictions for Haumea's ring assuming I/F = 0.09. The first
           part shows the estimates (in mJy) for our Spitzer and Herschel measurements, the second
           part includes calculations (in $\mu$Jy!) for JWST-MIRI filters. During the 2005/2007
           Spitzer observing epochs the ring contribution in the 24-$\mu$m band is expected
           to be below 1\,$\mu$Jy and therefore negligible. The MIRI imager
           sensitivities (SNR = 10 in 10,000 sec) are increasing from about 1\,$\mu$Jy at
           15\,$\mu$m to about 10\,$\mu$Jy at 25.5\,$\mu$m.} \label{tbl:ringflux}
\begin{tabular}{lrccc}
    \noalign{\smallskip}
    \hline
    \noalign{\smallskip}
    Observation     & 70\,$\mu$m & 100\,$\mu$m & 160\,$\mu$m & 250/350\,$\mu$m \\
    epoch           & [mJy]      & [mJy]       & [mJy]       & [mJy]     \\
    \noalign{\smallskip}
    \hline
    \noalign{\smallskip}
    Jun 2005 &  $<$0.05    & -    & -   & -/- \\
    Jul 2007 &  0.15       & -    & -   & -/- \\
    Dec 2009 &  0.4        & 1.0  & 1.2 & -/- \\
    Jun 2010 &  0.6        & 1.4  & 1.4 & -/- \\
    Jan 2011 &  -          & -    & -   & 1.4/1.0 \\
    \noalign{\smallskip}
    \hline
    \hline
    \noalign{\smallskip}
    JWST-MIRI  & 15\,$\mu$m & 18\,$\mu$m & 21\,$\mu$m & 25.5\,$\mu$m \\
    prediction & [$\mu$Jy]  & [$\mu$Jy]  & [$\mu$Jy]  & [$\mu$Jy] \\
    \noalign{\smallskip}
    \hline
    \noalign{\smallskip}
    early 2022  & 0.012      & 0.28       & 2.5        & 23 \\
    mid 2025    & 0.023      & 0.48       & 4.0        & 35 \\
    \noalign{\smallskip}
    \hline
    \noalign{\smallskip}
\end{tabular}
\end{table}


In the work by \citet{Santos-Sanz2017}, the possible thermal contributions
of the satellites Hi'iaka and Namaka were estimated to be $\sim$6\% and 1.5\%, respectively,
but they did not consider these contributions in their radiometric analysis. And, the
ring was not known at that time. Here, we subtract the summed up minimum ring-satellite contribution
from the total observed Spitzer and Herschel fluxes:
0.002\,mJy and 1.1 (1.2)\,mJy for the MIPS 24- and 70-$\mu$m observations
in 2005 (2007); 1.6\,mJy, 2.6 (3.0)\,mJy, and 2.8 (3.0)\,mJy for the 
PACS 70, 100, and 160-$\mu$m observations in 2009 (2010); and, 2.5\,mJy 
and 1.7\,mJy for the SPIRE 250 and 350\,$\mu$m observations in 2011.
Depending on wavelengths and epoch, the Haumea-only fluxes are
about 5-23\% lower than the total observed fluxes for the most conservative
case (with the satellites
having an albedo above 0.7). In case of darker surfaces (albedo of 0.5 for
the satellites), the satellite/ring contributions would increase significantly
and could easily reach 1/3 of the observed total flux. However, in these cases,
the Haumea model would require a very high thermal inertia well above 10 to
lower it's flux contribution accordingly. At the same time, the amplitude of the thermal
lightcurve would also decrease, well below the observed
100 and 160\,$\mu$m amplitudes (see Figs.~\ref{fig:TPM100um2} and \ref{fig:obsmod}, bottom).

\section{Discussion}
\label{sec:discussion}

The size and shape estimate for Haumea and the ring discovery from the
multi-chord occultation observation were crucial for the reinterpretation
of the thermal emission measurements. From our calculations (Figs.~\ref{fig:TPM100um1}
and \ref{fig:TPM100um2}), it is very likely that the satellites and the ring
contributions to the measured signals are small. As a consequence, the
satellites (at least Hi'iaka) must have a high albedo (very likely above 0.5),
otherwise their thermal contribution would reach several milli-Jansky
which is problematic when trying to match the Haumea occultation size to
the ring-satellite-subtracted thermal flux. Haumea's thermal inertia
would be above 10\,J m$^{-2}$ s$^{-1/2}$ K$^{-1}$ (see also discussion below),
and the thermal lightcurve amplitude would shrink to 1-2\,mJy, well below the
observed amplitude.
The case of a dark Hi'iaka (and also Namaka) with p$_V$ $<$0.5 can therefore
be excluded. We should note here, that Namaka's flux is only about 20 to 25\%
of Hi'iaka's flux (for equal albedo), so excluding the case of a dark
Namaka is much more difficult.
The albedo limit of about 0.5 is not very well determined and it is related
to the assumption of a beaming parameter of $\eta$ = 1.0. A higher $\eta$-value
(e.g.\ $\eta$=1.3) for the satellites would lower the thermal emission contribution
by 10-20\% in the Herschel wavelength range (higher value for the shorter
wavelengths). In this case, a Hi'iaka albedo of $\sim$0.5 would still be compatible
with the observed system fluxes.
Additional small flux contributions could
also come from undiscovered small satellites, but previous search programs
\citep{Brown2006} excluded other satellites at the size of Namaka or larger.
However, the high albedo for the satellites is connected to small
diameters of about 300 and 150\,km for Hi'iaka and Namaka, respectively
(see Table~\ref{tbl:satflux} and the discussions in Section~\ref{sec:satsring}).
And since the masses are known, their densities must be well above 1.0\,g cm$^{-3}$
to comply with the mass and the brightness constraints. This makes
these satellites (or at least Hi'iaka) very special in the current size-density picture
(see \citealt{Kiss2018}). They show that all TNOs below 500\,km in diameter
seem to have densities well below 1.0\,g cm$^{-3}$, i.e.\ close to (or even below)
the density of pure water ice or granular ice with self-compression
\citep{Durham2005}. The indication for higher densities for Haumea satellites
points therefore to a formation process of these two satellites which is
different from other small-size TNOs. A density below 1.0\,g cm$^{-3}$ would also be the
expectation for Haumea's satellites if they have been formed by a collision.
Their density should represent the icy crust of Haumea which is expected to
be much lower than Haumea's bulk density of about 1.9\,g\,cm$^{-3}$. These
questions about Hi'iaka's and Namaka's density and their formation cannot
be answered conclusively by our thermal measurements, especially, since the masses
of both satellites are also uncertain \citep{Ragozzine2009}.

The two constraints from Figs.~\ref{fig:TPM100um1} and
\ref{fig:TPM100um2} indicate that Haumea's thermal inertia is very
likely between 2 and 10\,J m$^{-2}$ s$^{-1/2}$ K$^{-1}$ (possibly up to 15\,J m$^{-2}$ s$^{-1/2}$ K$^{-1}$ if the
a-axis is larger than the default value). This can be tested on
all thermal data combined with a standard radiometric technique.
In a first case (case I) we do not subtract any ring or satellite
contribution. In case II and case III, we use all MIPS/PACS/SPIRE
fluxes with the corresponding satellite (see Table~\ref{tbl:satflux})
and ring contributions (see Table~\ref{tbl:ringflux}) subtracted.
In case II, we assume a high albedo of p$_V$=0.8 for
both satellites (low satellite flux contribution), in case III
we assume p$_V$=0.5 (producing higher satellite flux contributions).
Then we search for the thermal inertia that produces the best fit
to all data simultaneously, without violating the occultation 
and H-mag constraints.

\begin{table*}[h!tb]
  \centering
  \caption{Radiometric analysis of all thermal measurements combined. Cases are explained in the text.
           We accepted values for the reduced $\chi^{2}$ up to 1.4, with the optimum $\chi^{2}$ around
           0.7 (case I \& II) and 0.8 (case III).}
  \label{tbl:tpm}
  \begin{tabular}{rrl}
    \noalign{\smallskip}
    \hline \hline
    \noalign{\smallskip}
         &  TI range  &          \\
    Case & [J m$^{−2}$ s$^{−1/2}$ K$^{−1}$] & comments \\
    \noalign{\smallskip}
    \hline
    \noalign{\smallskip}
      I & 1-3  & trend in obs/mod ratios with wavelengths, \\
        &      & conflict with mass-brightness constraints on the satellites  \\
     II & 2-7  & best solution at TI=5 ($\chi^{2}_{r}$=0.7) \\ 
        &      & compatible with satellite mass/density estimates \\
        &      & match to absolute flux level and lightcurve amplitude \\
    III & 7-11 & amplitude fit and overall match to flux levels degrading, \\
        &      & trend in obs/mod ratios with wavelengths \\
    \noalign{\smallskip}
    \hline
    \noalign{\smallskip}
  \end{tabular}
\end{table*}

\begin{figure*}
  \centering
  \includegraphics[angle=90,width=0.67\hsize]{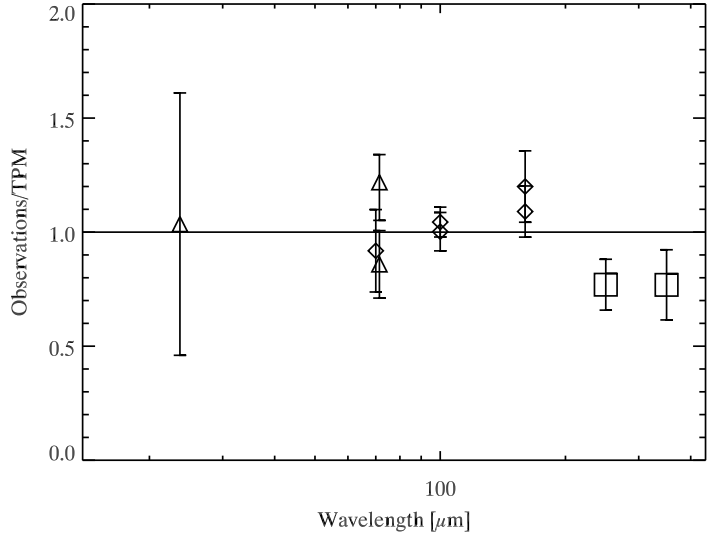}
  \includegraphics[angle=90,width=0.67\hsize]{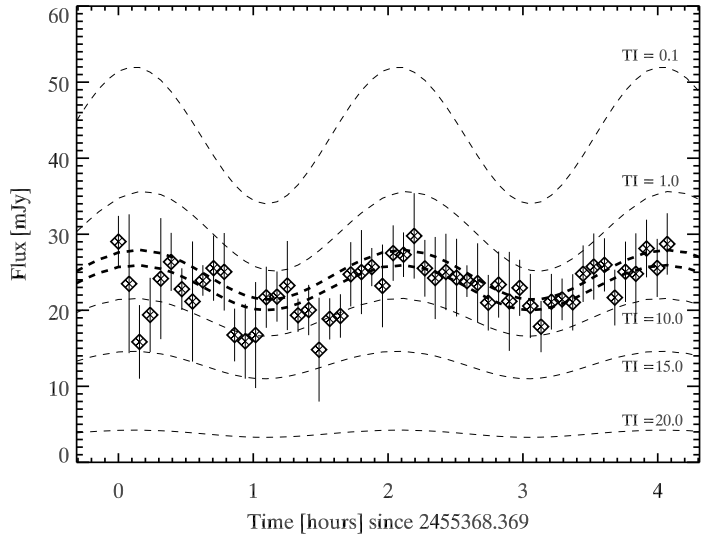}
  \caption{Top: All observations (MIPS and PACS lightcurves have been averaged),
           divided by the corresponding TPM predictions
           ($\Gamma$ = 5\,J m$^{-2}$ s$^{-1/2}$ K$^{-1}$, r.m.s.\ of surface slopes
           of 0.2, q= 0.65, occultation-derived 3-D shape and spin properties for pole 1,
           Haumea emissivity of 0.9 in the PACS range and 0.8 in the SPIRE range).
           Triangles are Spitzer-MIPS data, diamonds are Herschel-PACS, and boxes are
           Herschel-SPIRE observations. Small thermal contributions (case II)
           from the ring and the satellites were subtracted from the observed fluxes. The 24\,$\mu$m data point
           is a non- or marginal detection, the 70, 100, and 160\,$\mu$m points are
           well matched. At 250 and 350\,$\mu$m we might have overestimated the
           ring and satellite contributions (see text). Note that the true size, shape and
           spin properties have been used for the ratio calculations.
           Bottom: the most-accurate 100\,$\mu$m lightcurve
           (PACS, second epoch in Jun 2010), with minimum ring/satellite contributions
           subtracted (case II), shown together with absolute model predictions over-plotted.
           The two thicker dashed lines are related to $\Gamma$ = 4\,J m$^{-2}$ s$^{-1/2}$ K$^{-1}$
           (higher one) and $\Gamma$ = 6\,J m$^{-2}$ s$^{-1/2}$ K$^{-1}$ (lower one).}
  \label{fig:obsmod}
\end{figure*}

We summarized our results in Table~\ref{tbl:tpm}. The $\chi^2$ minima
are not changing much for the three cases, but when studying the fits to the different thermal
lightcurves individually, we see problems for low ($<$3\,J m$^{-2}$ s$^{-1/2}$ K$^{-1}$) and
high ($\Gamma$ $>$ 8\,J m$^{-2}$ s$^{-1/2}$ K$^{-1}$) thermal inertias. These problems
are nicely visible in Figure~\ref{fig:obsmod} (bottom part) where the different solutions
for the highest-quality PACS 100\,$\mu$m lightcurve are shown. Case I and case III also fail to
match the lightcurve-averaged fluxes in different bands (70, 100, 160\,$\mu$m) equally well:
In the observation-to-model ratios one can see a strong trend with wavelengths (the models
overestimate the 70\,$\mu$m fluxes and underestimate the 160\,$\mu$m levels).
Another argument against case I: The satellite mass estimates by \citet{Ragozzine2009} lead
to a thermal signal of at least 1-2\,mJy in the PACS bands for the highest possible
albedo and significantly more for a lower albedo (and/or in case of more, so-far undiscovered,
satellites). The case I solution can therefore only produce a lower limit for 
Haumea's thermal inertia. Case III is also very problematic: if the ring and the satellites contribute too much, then
Haumea's thermal inertia has to increase accordingly and this would lower
the thermal lightcurve amplitude well below the observed level. The offset
in observation-to-model ratio at sub-millimeter range would also increase and be more
difficult to explain (see lower part of Figure~\ref{fig:obsmod}).

The overall best solution (in reduced $\chi^2$, and also when inspecting the match to the absolute
fluxes and thermal lightcurve), is found for case II, with TPM settings of $\Gamma$ $\sim$5,
a phase integral of 0.65, and a surface roughness of $\sim$0.2 r.m.s.\ of surface slopes.
Figure~\ref{fig:obsmod} shows our best match to the data for case II, with only
the small ring/satellite contributions (5-20\% to the total observed fluxes) subtracted.
In the upper part of the figure we have calculated Haumea's thermal emission at the
given observing geometry for each thermal measurement, and we show the
observations-to-model ratios as a function of wavelengths.
At 70, 100, and 160\,$\mu$m, the measurements are well matched, at 24\,$\mu$m
the model prediction is in agreement with the non- or marginal detection,
but at 250 and 350\,$\mu$m the predictions overestimate the fluxes. It seems that
the ring and satellite contributions are negligible at these wavelengths,
indicating that their sub-millimeter emissivity is lower than the assumed value of $\epsilon$ = 0.9
(for the satellites) and 1.0 (for the ring).
Another option is a higher I/F value for the ring: with increasing albedo the thermal
ring emission is going down and for an I/F of 0.5 it would decrease to about half
the assumed value, but this would also lower the ring contribution in the PACS range
and it would only partly explain the SPIRE data problem. But both effects combined
would lower the sub-millimeter thermal contribution of the ring and the satellites and would
bring the observation-to-model ratios at 250 and 350\,$\mu$m in Figure~\ref{fig:obsmod}
closer to 1.0.
Our current ring model does not depend on particle sizes and it is therefore not
possible to constrain the particle sizes. However, a recent study on the particle dynamics
of Haumea's dust ring \citep{Kovacs2018} found that particles with sizes of $\sim$1\,$\mu$m
would accumulate circularly in a narrow ring near the 3:1 spin-orbit resonance and
survive for a reasonable time periods. Such small grain sizes would make the ring
more transparent at far-IR wavelengths and lead to a better match between our
Haumea predictions and the (case II) fluxes.

How does Haumea's thermal inertia compare with other distant TNOs or
with large dwarf planets with known thermal properties?
\citet{Lellouch2013} analyzed a large sample of TNOs and found a
$\Gamma$ = 2.5 $\pm$ 0.5\,J m$^{-2}$ s$^{-1/2}$ K$^{-1}$ for objects
at heliocentric distances of r$_{helio}$ = 20--50\,AU. Haumea is at r$_{helio}$ = $\sim$51\,AU
and the thermal inertia should therefore be slightly lower due to the lower
surface temperatures: assuming that the T$^3$ term dominates in the thermal
conductivity (the thermal inertia scales with $\sim$ r$^{−3/4}$; see e.g. \citealt{Delbo2015}).
On the other hand, the thermal inertia of Pluto and Charon are larger:
   $\Gamma_{Pluto}$ = 16-26\,J m$^{-2}$ s$^{-1/2}$ K$^{-1}$ and
   $\Gamma_{Charon}$ = 9-14\,J m$^{-2}$ s$^{-1/2}$ K$^{-1}$ \citep{Lellouch2011,Lellouch2016}.
A Pluto(Charon)-like surface has then a thermal inertia of 11.3-18.3(6.3-9.9)\,J m$^{-2}$ s$^{-1/2}$ K$^{-1}$
at the distance of Haumea (about a factor of 1.4 lower). But the high $\Gamma$-values
for Pluto and Charon are thought to be caused by large diurnal skin depth due to their
slow rotation ($\sim$ P$^{1/2}$ dependence; see also discussion in \citealt{Kiss2018}).
Haumea rotates much faster than Pluto and Charon (6.38 d orbital/rotation period of
Pluto-Charon versus 3.91\,h for Haumea), and also faster than typical TNOs
(P = $\sim$6-12\,h; \citealt{Duffard2009, Thirouin2016}). Taking the skin-depth effect into account would lower the
expected thermal inertia to about 1.8-2.9(1.0-1.5)\,J m$^{-2}$ s$^{-1/2}$ K$^{-1}$
for Haumea, if we assume a Pluto(Charon)-like surface. Our derived values
for Haumea are a factor of 2-3 higher and point to a more compact (hence higher
conductivity) surface than compared to Pluto or Charon.

The phase integral q is a critical property in the thermophysical model concept for
high-albedo objects like Haumea. The thermal flux depends mainly on the object's size
and its Bond albedo A$_{b}$, with A$_{b}$ = p$_V$ $\cdot$ q.
For objects with a standard scattering behavior in the visible, the phase slope
G is given with 0.15, leading to a phase integral of q = 0.39 \citep{Bowell1989}.
This works well for low-albedo objects, but fails for high-albedo objects.
Pluto's phase integral was found to be q = 0.8 \citep{Lellouch2000}, which was used by \citet{Stansberry2008}
for the interpretation of the Spitzer measurements of Haumea. \citet{Lellouch2010}
adopted a phase integral q = 0.7, intermediate between those estimated for
Pluto (0.8) and Charon (0.6) \citep{Lellouch2000}. \citet{Santos-Sanz2017} determined a high
phase integral ($>$0.73) for Haumea by matching an overestimated 100/160-$\mu$m lightcurve
amplitude.
Using the phase integral formula by \citet{Brucker2009}: q = 0.336 $\cdot$ p$_V$ + 0.479,
we find q = 0.65 (for p$_V$ = 0.51). We used different values for q, but there are
simply too many unknowns in the system to constrain its value. Overall, we find better
fits to the data for higher values (q=0.65 to 0.8), the lower values (q=0.45) would 
push Haumea's thermal inertia to larger values (above 10\,J m$^{-2}$ s$^{-1/2}$ K$^{-1}$)
and decrease the thermal lightcurve amplitudes too much. However, the combination of a
low q-values (around 0.5) with a smooth surface (r.m.s. $<$ 0.2) would also lead to
an acceptable solution with $\Gamma$ = 5\,J m$^{-2}$ s$^{-1/2}$ K$^{-1}$, and in agreement
with the observed flux and amplitude levels (see Fig.~\ref{fig:TPM100um1}).

The study of the DRS via thermal data is very uncertain. In Figure~\ref{fig:PacsLcData} it
is difficult to identify which of the 100\,$\mu$m maxima is higher, the errors are large,
and the outliers at 40\,mJy are difficult to judge. Following the DRS phasing in 
\citet{Santos-Sanz2017} and also in \citet{Gourgeot2016}, the DRS is 
connected to the second maximum (at rotational phases around 0.6 in Fig.~\ref{fig:PacsLcData}
or at 2.15 hours in Fig.~\ref{fig:obsmod}, lower part).
\citet{Lacerda2008} suggested three different DRS models with a darker region covering only a few percent of 
the maximum equatorial cross-sectional area, up to a model where a whole hemisphere has
a darker albedo. A few percent change in albedo between the DRS and the rest of the surface
can easily explain the two maxima in reflected light, with the DRS connected to the
lower optical maximum, hence the higher thermal maximum.
At Herschel/Spitzer wavelengths, also the thermal inertia plays a role, and if there is
a variation in thermal inertia, this could either compensate an albedo effect or enhance
it, and also surface roughness and phase integral play a role. But such an investigation
would require a high-quality thermal lightcurve, close in time with an optical ligthcurve
to constrain the physical and thermal properties of the DRS region.

\section{Conclusions and outlook}
\label{sec:conclusion}

The previously published thermal lightcurves \citep{Lellouch2010,Santos-Sanz2017}
showed a very large amplitude which led to a misinterpretation of Haumea's thermal
properties and systematically underestimated (by $\sim$25\%) radiometric sizes.
We have reanalyzed the Spitzer and Herschel far-infrared and sub-millimeter
observations from 2005 to 2011 with the latest reduction and calibration
schemes, and a more elaborate sky background handling.
The new multi-filter Herschel-PACS
fluxes and thermal lightcurves are presented in Tables~\ref{app:lc100a},
\ref{app:lc100b}, \ref{app:lc160}, \ref{app:phot}. The corresponding background-eliminated
Herschel-PACS maps are available from the HSA. All thermal observations are measurements of
the combined Haumea-ring-satellite system and their interpretation requires
to consider possible ring and satellite contributions. Only the thermal lightcurve
amplitude (measured at 70, 100, and 160\,$\mu$m) is connected to Haumea itself.
We confirm the 3.91\,h periodicity of the thermal lightcurve and we see that
thermal and visual lightcurves are in phase.

We estimated the time-dependent ring fluxes by using a simple thermal model from \citet{Lellouch2017},
based on ring properties of Saturn and Chariklo. During the Spitzer observations in
2005/2006 the ring emission was small (and almost negligible), but due to the
opening of the ring plane, the ring contributed $\sim$1-1.5\,mJy to the measured
Herschel signals in 2009-2011.


The mass and brightness estimates \citep{Brown2006, Ragozzine2009} for the
two Haumea satellites Hi'iaka and Namaka are crucial for the interpretation
of the thermal measurements. We calculated the far-IR fluxes for 
both satellites for a size-albedo range which is compatible with these
constraints and under the assumption of a density range between 0.5 and 1.5\,g cm$^{-3}$.
In case of a high density of 1.5\,g cm$^{-3}$ (requiring a geometric albedo of $\sim$0.8),
the satellites add about 1-1.5\,mJy to the observed flux. More moderate
density assumptions (related to a lower albedo) lead to much higher satellite
fluxes and cause severe problems in interpreting the remaining Haumea-only far-IR fluxes.

The key element for the re-interpretation of the Haumea measurements, however,
is coming from a successful occultation measurement \citep{Ortiz2017}.
It led to the discovery of the ring around Haumea, a highly-accurate 
ellipse fit to the multi-chord event (which happened very close to the
lightcurve minimum), and allowed to reconstruct the objects size, shape,
and spin-properties. We used the occultation-derived properties to take
a closer look at the possible thermal and physical properties of Haumea
and it's two satellites. We find that Haumea has a thermal inertia
of about 5\,J m$^{-2}$ s$^{-1/2}$ K$^{-1}$ (in combination with a surface
roughness r.m.s.\ = 0.2, a phase integral q = 0.65, and the pole~1
solution in \citealt{Ortiz2017}).
A much smaller of the thermal inertia value would require
that the ring and the satellite fluxes are completely negligible, a
higher value would lower the thermal lightcurve amplitude below the 
observed values. A lower phase integral (q = 0.5) would push the 
solution to higher thermal inertia (up to about 10\,J m$^{-2}$ s$^{-1/2}$ K$^{-1}$),
assuming a similar surface roughness.
When the degree of surface roughness is varied from
0.0 (smooth) to 0.5 (see \citealt{Fornasier2013}) the derived value
of Haumea's thermal inertia would also vary in the range
between 3 and 10\,J m$^{-2}$ s$^{-1/2}$ K$^{-1}$, with the smooth
surface connected to a lower thermal inertia value. The most
extreme thermal inertia of about $\sim$15\,J m$^{-2}$ s$^{-1/2}$ K$^{-1}$
is found for a low phase integral (q$<$0.5) combined with a high
surface roughness (r.m.s.$\gtrsim$0.5).

The satellites are very likely in the smaller size regime
($\sim$300\,km diameter for Hi'iaka and $\sim$150\,km for Namaka) linked to
a high albedo p$_{V}$ $>>$0.5, and connected to an unexpectedly high
density $>$1\,g cm$^{-1}$. Lower albedo solutions would push their
thermal emission to large values which are not compatible with the
occultation-derived size-shape-spin and our radiometric thermal-inertia
solution. Our fit to the sub-millimeter thermal measurements improves if
we lower the emissivity of the ring ($\epsilon$ = 1.0 assumed) and the
satellites for which we assumed $\epsilon$ = 0.9.

Haumea's ring contribution to the total thermal flux will increase
significantly over the next decades (Figure~\ref{fig:ringpred}).
The JWST-MIRI imager sensitivities (about 1\,$\mu$Jy at 15\,$\mu$m
to about 10\,$\mu$Jy at 25.5\,$\mu$m for a SNR of 10 in 10,000\,sec)
would be perfectly suitable to measure the ring-only fluxes, but
the spatial resolution will not be sufficient to separate the ring
from the main body. However, MIRI measurements would allow us to
confirm Haumea's thermal properties from the thermal lightcurve
amplitude and the shape of its spectral energy distribution. Then,
in comparison with the Spitzer and Herschel data, the MIRI multi-filter
data will provide a higher-quality characterization
of the ring properties as a byproduct.

Measurements with JWST-MIRI will be very sensitive to the thermal properties of Haumea.
Based on our best-fit TPM ($\Gamma$=5\,J m$^{-2}$ s$^{-1/2}$ K$^{-1}$, q=0.65, r.m.s.\ = 0.2)
for Haumea we predicted fluxes in three long-wavelengths
MIRI channels (see Table~\ref{tbl:haumea_miri}).

\begin{table*}[h!tb]
 \caption{JWST-MIRI predictions for Feb 1, 2022 at r$_{helio}$ = 50.18\,au, $\Delta$ = 50.00\,au,
  $\alpha$= 1.1$^{\circ}$, at a solar elongation of 100$^{\circ}$ when Haumea is visible for JWST.
  The first column in each band gives the average flux level, the second the lightcurve variations.
  Note that at shorter wavelengths below $\sim$20\,$\mu$m there is still a significant contribution
  from reflected light in MIRI bands. These predictions depend also on the phase integral (here: q = 0.65)
  and the surface roughness (here: r.m.s.\ = 0.2) and have therefor significant uncertainties.}
 \label{tbl:haumea_miri}
\begin{tabular}{rrrrr}
\noalign{\smallskip}
\hline \hline
\noalign{\smallskip}
Thermal inertia            & \multicolumn{2}{c}{21.0\,$\mu$m}  & \multicolumn{2}{c}{25.5\,$\mu$m} \\
$\Gamma$ [J m$^{−2}$ s$^{−1/2}$ K$^{−1}$]   & flx[$\mu$Jy] & amp[$\mu$Jy] & flx[$\mu$Jy] & amp[$\mu$Jy] \\
\noalign{\smallskip}
\hline
\noalign{\smallskip}
0.5 &  33 & $\pm$6 & 230 & $\pm$ 46 \\
2.0 &  21 & $\pm$3 & 131 & $\pm$ 12 \\
5.0 &  19 & $\pm$3 & 110 & $\pm$ 10 \\
10  &  10 & $\pm$2 &  57 & $\pm$ 10 \\
\noalign{\smallskip}
\hline
\end{tabular}
\end{table*}

The MIRI imager sensitivities (SNR = 10 in 10,000 sec) are increasing from about
1\,$\mu$Jy at 15\,$\mu$m to about 10\,$\mu$Jy at 25.5\,$\mu$m. It will therefore be
possible to obtain SNR well above 25 in 1000\,sec in the 25.5\,$\mu$m filter at 
all rotational phases. A second, shorter-wavelength band (possibly at 21\,$\mu$m)
will allow characterizing the spectral slope and help to disentangle the
ring and satellite contributions. At shorter wavelength below $\sim$20\,$\mu$m
there are still strong contributions from reflected Sun light.

\section*{Acknowledgement}
The research leading to these results has received funding from the European
Union's Horizon 2020 Research and Innovation Programme, under Grant Agreement no 687378.
Funding from the Spanish grant AYA-2017-89637-R is acknowledged.

\appendix

\section{Herschel-PACS photometric flux densities}
\label{app:pacs}

We performed aperture photometry on the reduced, calibrated,
and background eliminated final images. The fluxes were aperture
corrected, color-corrected (correction factors 1.00, 0.98, and
0.99 at 70.0, 100.0, and 160.0\,$\mu$m, respectively). We also
included the recommended 5\% absolute flux error when we
use PACS fluxes in radiometric calculations.

\subsection{Lightcurve measurements at 100\,$\mu$m}

At 100\,$\mu$m, we merged always three repetitions (shifted 
by 1) before extracting the aperture photometry (first line:
repetitions 1-3, second line 2-4, third line 3-5, etc.).


\begin{table}[h!tb]
 \caption{Herschel-PACS lightcurve observations (OBSIDs 1342188470, 1342188520) at 100\,$\mu$m on 2009 Dec 23. The
          times are observation mid-times in the Herschel reference frame.}
 \label{app:lc100a}
\begin{tabular}{lrrrr}
\noalign{\smallskip}
\hline \hline
\noalign{\smallskip}
mid-time   & \multicolumn{2}{c}{in-band flux \& error} & \multicolumn{2}{c}{abs. flux \& error} \\
obs. epoch & [mJy] &  [mJy] & [mJy] &  [mJy] \\
\noalign{\smallskip}
\hline
\noalign{\smallskip}
2455188.7492 & 25.49 &  9.37 & 26.01 &  9.46 \\
2455188.7527 & 25.06 &  9.04 & 25.57 &  9.12 \\
2455188.7562 & 23.55 &  8.96 & 24.03 &  9.04 \\
2455188.7597 & 29.47 & 10.68 & 30.07 & 10.78 \\
2455188.7632 & 34.11 &  7.85 & 34.81 &  8.03 \\
2455188.7667 & 34.81 &  6.02 & 35.52 &  6.27 \\
2455188.7702 & 32.41 &  6.33 & 33.07 &  6.53 \\
2455188.7737 & 31.02 &  7.81 & 31.65 &  7.96 \\
2455188.7772 & 26.47 &  8.16 & 27.01 &  8.27 \\
2455188.7807 & 23.51 &  6.91 & 23.99 &  7.01 \\
2455188.7842 & 18.59 &  4.91 & 18.97 &  5.00 \\
2455188.7877 & 18.91 &  5.97 & 19.30 &  6.05 \\
2455188.7912 & 17.76 &  6.18 & 18.12 &  6.24 \\
2455188.7947 & 18.16 &  6.90 & 18.54 &  6.96 \\
2455188.7982 & 23.24 &  5.33 & 23.72 &  5.46 \\
2455188.8017 & 24.53 &  5.53 & 25.03 &  5.67 \\
2455188.8052 & 25.69 &  5.71 & 26.22 &  5.86 \\
2455188.8087 & 20.00 &  5.84 & 20.40 &  5.92 \\
2455188.8122 & 21.57 &  5.84 & 22.01 &  5.94 \\
2455188.8156 & 18.31 &  6.77 & 18.69 &  6.83 \\
2455188.8191 & 22.52 &  4.71 & 22.98 &  4.84 \\
2455188.8226 & 23.10 &  4.88 & 23.57 &  5.02 \\
2455188.8261 & 28.57 &  4.01 & 29.15 &  4.26 \\
2455188.8296 & 26.15 &  5.29 & 26.69 &  5.45 \\
2455188.8331 & 23.01 &  5.68 & 23.48 &  5.80 \\
2455188.8366 & 21.25 &  5.29 & 21.68 &  5.39 \\
2455188.8401 & 24.08 &  5.72 & 24.57 &  5.84 \\
2455188.8436 & 40.97 &  3.67 & 41.80 &  4.21 \\
2455188.8471 & 42.61 &  5.39 & 43.48 &  5.79 \\
2455188.8506 & 44.43 &  3.78 & 45.34 &  4.38 \\
2455188.8541 & 26.61 &  5.49 & 27.15 &  5.65 \\
2455188.8576 & 23.39 &  4.83 & 23.86 &  4.97 \\
2455188.8611 & 18.93 &  4.99 & 19.32 &  5.08 \\
2455188.8646 & 22.14 &  4.99 & 22.59 &  5.11 \\
2455188.8681 & 25.24 &  5.06 & 25.76 &  5.22 \\
2455188.8716 & 27.79 &  5.84 & 28.35 &  6.01 \\
2455188.8751 & 20.83 &  6.74 & 21.26 &  6.82 \\
2455188.8784 & 20.01 &  6.40 & 20.42 &  6.48 \\
\noalign{\smallskip}
\hline
\end{tabular}
\end{table}

\begin{table}[h!tb]
 \caption{Herschel-PACS lightcurve observations (OBSIDs 1342198851, 1342198905/906)
          at 100\,$\mu$m on 2010 Jun 20/21. The
          times are observation mid-times in the Herschel reference frame.}
 \label{app:lc100b}
\begin{tabular}{lrrrr}
\noalign{\smallskip}
\hline \hline
\noalign{\smallskip}
mid-time   & \multicolumn{2}{c}{in-band flux \& error} & \multicolumn{2}{c}{abs. flux \& error} \\
obs. epoch & [mJy] &  [mJy] & [mJy] &  [mJy] \\
\noalign{\smallskip}
\hline
\noalign{\smallskip}
2455368.3690 & 31.39 &  2.99 & 32.03 &  3.38 \\
2455368.3723 & 25.96 &  9.04 & 26.49 &  9.13 \\
2455368.3755 & 18.46 &  4.75 & 18.84 &  4.84 \\
2455368.3788 & 21.94 &  4.76 & 22.39 &  4.88 \\
2455368.3821 & 26.62 &  7.85 & 27.16 &  7.96 \\
2455368.3853 & 28.77 &  3.54 & 29.36 &  3.82 \\
2455368.3886 & 25.27 &  2.44 & 25.78 &  2.74 \\
2455368.3919 & 23.67 &  7.81 & 24.15 &  7.90 \\
2455368.3951 & 26.38 &  1.52 & 26.92 &  2.01 \\
2455368.3984 & 27.97 &  4.05 & 28.54 &  4.29 \\
2455368.4017 & 27.50 &  4.91 & 28.06 &  5.10 \\
2455368.4049 & 19.36 &  3.33 & 19.75 &  3.47 \\
2455368.4082 & 18.54 &  4.83 & 18.92 &  4.92 \\
2455368.4114 & 19.34 &  6.90 & 19.74 &  6.97 \\
2455368.4147 & 24.22 &  3.83 & 24.71 &  4.02 \\
2455368.4180 & 24.30 &  3.07 & 24.80 &  3.30 \\
2455368.4212 & 25.73 &  5.71 & 26.25 &  5.86 \\
2455368.4245 & 21.89 &  1.90 & 22.34 &  2.19 \\
2455368.4278 & 22.57 &  3.11 & 23.03 &  3.31 \\
2455368.4310 & 17.46 &  6.77 & 17.82 &  6.83 \\
2455368.4343 & 21.44 &  2.49 & 21.87 &  2.71 \\
2455368.4376 & 21.81 &  2.65 & 22.25 &  2.86 \\
2455368.4408 & 27.18 &  4.01 & 27.73 &  4.23 \\
2455368.4441 & 27.46 &  5.36 & 28.02 &  5.53 \\
2455368.4473 & 28.11 &  2.08 & 28.68 &  2.51 \\
2455368.4506 & 25.68 &  5.29 & 26.20 &  5.44 \\
2455368.4539 & 29.91 &  3.33 & 30.52 &  3.65 \\
2455368.4571 & 29.70 &  2.52 & 30.31 &  2.93 \\
2455368.4604 & 32.12 &  5.39 & 32.78 &  5.62 \\
2455368.4637 & 27.95 &  3.46 & 28.52 &  3.73 \\
2455368.4669 & 26.69 &  5.23 & 27.24 &  5.40 \\
2455368.4702 & 27.48 &  4.83 & 28.04 &  5.02 \\
2455368.4735 & 26.75 &  4.93 & 27.29 &  5.11 \\
2455368.4767 & 26.30 &  1.61 & 26.83 &  2.08 \\
2455368.4800 & 26.05 &  5.06 & 26.58 &  5.23 \\
2455368.4833 & 23.51 &  3.47 & 23.99 &  3.66 \\
2455368.4865 & 25.86 &  4.15 & 26.38 &  4.35 \\
2455368.4898 & 23.69 &  6.40 & 24.17 &  6.51 \\
2455368.4930 & 25.44 &  3.47 & 25.95 &  3.69 \\
2455368.4963 & 23.09 &  4.06 & 23.56 &  4.22 \\
2455368.4996 & 20.43 &  3.21 & 20.85 &  3.37 \\
2455368.5028 & 23.64 &  3.47 & 24.12 &  3.66 \\
2455368.5061 & 23.94 &  2.46 & 24.43 &  2.74 \\
2455368.5094 & 23.56 &  3.51 & 24.04 &  3.70 \\
2455368.5126 & 27.26 &  3.47 & 27.81 &  3.72 \\
2455368.5159 & 28.18 &  4.11 & 28.75 &  4.35 \\
2455368.5192 & 28.39 &  3.20 & 28.97 &  3.50 \\
2455368.5224 & 24.18 &  3.47 & 24.67 &  3.67 \\
2455368.5257 & 27.50 &  3.73 & 28.06 &  3.97 \\
2455368.5289 & 27.24 &  5.14 & 27.80 &  5.32 \\
2455368.5322 & 30.50 &  3.47 & 31.12 &  3.79 \\
2455368.5355 & 28.00 &  3.55 & 28.57 &  3.82 \\
2455368.5386 & 31.09 &  3.73 & 31.72 &  4.04 \\
\noalign{\smallskip}
\hline
\end{tabular}
\end{table}

\subsection{Lightcurve measurements at 160\,$\mu$m}

At 160\,$\mu$m, we merged always six repetitions (shifted
by 3) before extracting the aperture photometry (first line:
repetitions 1-6, second line 4-9, third line 7-12, etc.).

\begin{table}[h!tb]
 \caption{Herschel-PACS lightcurve observations at 160\,$\mu$m on 2009 Dec 23
          (OBSIDs 1342188470, 1342188520) and 2010 Jun 20/21 (OBSIDs 1342198851, 1342198905/906).
          The times are observation mid-times in the Herschel reference frame.}
 \label{app:lc160}
\begin{tabular}{lrrrr}
\noalign{\smallskip}
\hline \hline
\noalign{\smallskip}
mid-time   & \multicolumn{2}{c}{in-band flux \& error} & \multicolumn{2}{c}{abs. flux \& error} \\
obs. epoch & [mJy] &  [mJy] & [mJy] &  [mJy] \\
\noalign{\smallskip}
\hline
\noalign{\smallskip}
2455188.7545 & 24.76 & 12.66 & 25.01 & 12.72 \\
2455188.7650 & 30.12 & 15.51 & 30.42 & 15.58 \\
2455188.7755 & 23.75 & 15.33 & 23.99 & 15.37 \\
2455188.7859 & 18.27 & 14.13 & 18.45 & 14.16 \\
2455188.7964 & 24.45 & 14.94 & 24.69 & 14.99 \\
2455188.8069 & 21.03 & 15.99 & 21.24 & 16.02 \\
2455188.8174 & 18.72 & 15.44 & 18.91 & 15.47 \\
2455188.8279 & 27.61 & 15.00 & 27.89 & 15.06 \\
2455188.8384 & 35.60 & 13.82 & 35.96 & 13.94 \\
2455188.8489 & 31.83 & 12.85 & 32.16 & 12.95 \\
2455188.8593 & 34.21 & 11.94 & 34.55 & 12.06 \\
2455188.8698 & 32.95 & 12.02 & 33.28 & 12.13 \\
\noalign{\smallskip}
2455368.3739 & 37.54 &  7.44 & 37.92 &  7.67 \\
2455368.3837 & 31.83 &  8.60 & 32.15 &  8.75 \\
2455368.3935 & 34.26 &  7.23 & 34.61 &  7.43 \\
2455368.4033 & 40.41 &  4.82 & 40.82 &  5.23 \\
2455368.4131 & 35.11 &  7.58 & 35.46 &  7.78 \\
2455368.4229 & 28.59 &  7.91 & 28.88 &  8.03 \\
2455368.4327 & 24.13 &  8.30 & 24.37 &  8.39 \\
2455368.4425 & 24.37 &  7.48 & 24.62 &  7.58 \\
2455368.4522 & 22.05 &  8.87 & 22.27 &  8.94 \\
2455368.4620 & 32.03 &  6.84 & 32.35 &  7.03 \\
2455368.4718 & 33.66 &  7.97 & 34.00 &  8.15 \\
2455368.4816 & 21.51 & 11.32 & 21.73 & 11.37 \\
2455368.4914 & 29.50 & 10.45 & 29.79 & 10.55 \\
2455368.5012 & 27.60 &  8.83 & 27.88 &  8.94 \\
2455368.5110 & 20.43 &  9.93 & 20.64 &  9.99 \\
2455368.5208 & 27.05 &  9.48 & 27.32 &  9.57 \\
2455368.5306 & 20.66 &  8.31 & 20.87 &  8.37 \\
\noalign{\smallskip}
\hline
\end{tabular}
\end{table}

\subsection{Multi-filter measurements}

\begin{table*}[h!tb]
 \caption{Herschel-PACS 3-band observations from 2010 Jun 21. The
          times are observation mid-times in the Herschel reference frame.}
 \label{app:phot}
\begin{tabular}{llrrrrrl}
\noalign{\smallskip}
\hline \hline
\noalign{\smallskip}
mid-time   &      & \multicolumn{2}{c}{in-band flux \& error} & $\lambda_{ref}$ & \multicolumn{2}{c}{abs. flux \& error} & \\
obs. epoch & band & [mJy] &  [mJy] & [$\mu$m] & [mJy] &  [mJy] & OBSIDs \\
\noalign{\smallskip}
\hline
\noalign{\smallskip}
2455369.45202 & blue  & 13.79 & 2.30 &  70.0 & 13.8 & 2.4 & 1342198903/904 \\
2455369.46740 & green & 14.49 & 3.50 & 100.0 & 14.8 & 3.6 & 1342198905/906 \\
2455369.45933 & red   & 31.48 & 5.26 & 160.0 & 31.8 & 5.5 & 1342198903/904/905/906 \\
\noalign{\smallskip}
\hline
\end{tabular}
\end{table*}


\section*{References}
\bibliographystyle{elsarticle-harv}
\bibliography{AsteroidsGeneral}





\end{document}